\documentclass[12pt,preprint]{aastex}

\usepackage{ulem}
\usepackage[usenames]{color}

\newcommand{\Ni}{$^{56}$Ni}
\newcommand{\Co}{$^{56}$Co}
\newcommand{\Ms}{$M_{\odot}$}
\newcommand      \grays       {$\gamma$-rays}

\newcommand{\til}{~}

\shorttitle{Molecules in massive Pop. III supernovae}
\shortauthors{Cherchneff \& Dwek}

\begin{document}


\title{THE CHEMISTRY OF POPULATION III SUPERNOVA EJECTA: I - FORMATION OF MOLECULES IN THE EARLY UNIVERSE}


\author{Isabelle Cherchneff \altaffilmark{1} \& Eli Dwek\altaffilmark{2}}


\altaffiltext{1}{Department Physik, Universit{\"a}t Basel , CH-4056 Basel, Switzerland; isabelle.cherchneff@unibas.ch}
\altaffiltext{2}{Observational Cosmology Laboratory, Code 665, NASA Goddard Space Flight Center, Greenbelt, MD 20771, USA; eli.dwek@nasa.gov}

\begin{abstract}
We study the formation and destruction of molecules in the ejecta of Population~III supernovae (SNe) using a chemical kinetic approach to follow the evolution of molecular abundances from day 100 to day 1000 after explosion. The chemical species included in the study  range from simple di-atomic molecules to more complex dust precursor species. All relevant molecule formation and destruction processes that are unique to the SN environment are considered. Our work focuses on zero-metallicity progenitors with masses of 20, 170, and 270~\Ms, and we study the effect of different levels of heavy element mixing and the inward diffusion of hydrogen and helium on the ejecta chemistry.
We show that the ejecta chemistry does not reach a steady state within the relevant timespan ($\sim 3$~yr) for molecule formation, thus invalidating previous results relying on this assumption. The primary species formed in the harsh SN environment are O$_2$, CO, SiS, and SO. The SiO, formed as early as 200 days after explosion, is rapidly depleted by the formation of silica molecular precursors in the ejecta.
The rapid conversion of CO to C$_2$ and its thermal fractionation at temperatures above 5000 K allow for the formation of carbon chains in the oxygen-rich zone of the unmixed models, providing an important pathway for the formation of carbon dust in hot environments where the C/O ratio is less than 1. 
We show that the fully-mixed ejecta of a 170~\Ms\ progenitor synthesizes 11.3 \Ms\ of molecules whereas 20 \Ms\ and 270 \Ms\ progenitors produce 0.78, and 3.2~\Ms\ of molecules, respectively. The admixing of 10\% of hydrogen into the fully-mixed ejecta of the 170~\Ms\ progenitor increases its molecular yield to $\sim 47$~\Ms. The unmixed ejecta of a 170~\Ms\ progenitor supernova without hydrogen penetration synthesizes $\sim 37$~\Ms\ of molecules, whereas its 20~\Ms\ counterpart produces $\sim 1.2$~\Ms. This smaller efficiency at forming molecules is due to the large fraction of He$^+$ in the outer mass zone of the ejecta. Finally, we discuss the cosmological implication of molecule formation by Pop.~III SNe in the early universe.

\end{abstract}




\section{INTRODUCTION}

Large column densities of dust are required to explain the reddening of background quasars and Lyman $\alpha$ systems at high redshift (z $>$ 6) (Pettini et al. 1994, Pei \& Fall 1995) and about $ 2 \times 10^8$ \Ms\til of dust is derived from the infrared (IR) spectrum of the hyperluminous galaxy SDSS J1148+5251 at redshift z = 6.4 (Bertoldi et al. 2003, Robson et al. 2004, Beelen et al. 2006, Dwek et al. 2007). The origin of such large quantities of dust when the universe was less than 1 Gyr-old is still a matter of debate. In the local universe, dust forms in high density and temperature regions encountered in circumstellar environments such as the winds of Asymtotic Giant Branch (AGB) stars and  supergiant stars, the colliding winds of Wolf-Rayet stars, and finally, the ejecta of core-collapse supernovae (CCSNe). In our Galaxy, most of the dust is produced by low-mass stars ascending the AGB. However, their long evolutionary main-sequence lifetime (a few Gyrs) excludes them from being possible dust contributors at high redshift. Conversely, very massive stars evolve much more rapidly (time scales $\sim$ 1 Myr), and can be possible dust makers in the early universe. As to the first generation of stars, hereafter Pop~III stars, they are expected to be very massive (Omukai \& Nishi 1998, Abel et al 2002, Bromm et al. 2002). Indeed in the absence of metals, the cooling in primordial clouds is only provided by molecular hydrogen and thus precludes efficient gas fractionation. These Pop.~III stars firstly need to synthesize heavy elements by thermonuclear reactions in their cores and reach their explosive ends to possibly condense dust in their massive supernova ejecta. Therefore, pair-instability supernovae (PISNe) are perhaps the first dust contributors to the pristine, young universe. 

The build-up of a molecular phase is a pre-requisite to dust nucleation and condensation. Indeed, it provides the molecular precursors from which dust forms and its composition depends on the initial elemental composition of the gas and the physical processes pertaining to it. Furthermore, molecules produced in the early universe can have an important effect on the cooling of the interstellar medium. Molecules have been detected in low-redshift CCSNe as early as 100 days post-explosion. Specifically, the IR ro-vibrationals transitions of CO and SiO were detected in SN1987A (Catchpole \& Glass 1987, Meikle et al. 1989, Roche et al. 1991), CO fundamental bands were observed in the Type II SNe SN1995ad (Spyromilio \& Leibundgut 1996), SN1998s (Gerardy et al. 2000) and SN202dh (Pozzo et al. 2006) whilst SiO detection was reported by Kotak et al.~(2006) in SN2005af. More recently, detection of CO with the Spitzer satellite in Cas A, a 300 year-old supernova remnant, is reported by Rho et al. (2009). It is therefore reasonable to expect molecules to form in the ejecta of massive Pop.~III star supernovae. 

Existing models for CO and SiO formation in SN1987A ejecta only considered a limited number of chemical processes applicable to low-temperature gases and assumed that steady state held for chemistry in calculating the evolution of molecular masses (Petuchowski et al. 1989, Lepp et al. 1990, Liu \& Dalgarno, 1994, 1995, 1996, Clayton et al. 1999, 2001, Geahard et al. 1999). Models for the formation of dust in SN1987A and PISNe used a classical nucleation theory approach to describe the growth of solids in the ejecta, ignoring the nucleation stage of dust, in which gas phase species and dust molecular precursors are formed (Kozasa et al. 1989, Toddini \& Ferrara 2001, Nozawa et al. 2003 (hereafter NK03), Schneider et al. 2004 (hereafter SFS04)). Since the amount of carbon and oxygen locked up in CO is a major factor in determining the dust composition, Toddini \& Ferrrara (2001) and SFS04 do consider the formation of CO in fully microscopically-mixed ejecta. However, their treatment is extremely over-simplified as their CO chemistry includes only two chemical processes and is assumed once again at steady state. In a first attempt to model molecular formation with a chemical kinetic approach in fully microscopically-mixed Pop.~III SN ejecta, Cherchneff \& Lilly (2008) (herafter CL08) show that the chemistry does not reach a steady state over the timespan studied. Furthermore, they identified other chemical processes than those considered by SFS04 that are of paramount importance to the formation of CO and other molecular species in the ejecta. 

In this paper, we study the formation and evolution of molecules, including gas-phase dust precursors, in the ejecta of Pop~III SNe. We define a large set of chemical reactions relevant to the dense and hot SN environment but applicable to other circumstellar media as well. In addition to this extensive reaction network, with new updated rates compared to those used by CL08, we include processes unique to the radioactive environment of SN ejecta: destructive processed induced by Compton electrons created by the down-scattering of $\gamma$-rays, and by ultraviolet (UV) photons emitted by collisionally-excited atoms and ions. We consider fully microscopically-mixed and unmixed ejecta for different progenitor masses. In addition, the presence of hydrogen and helium can dramatically affect the chemistry of the ejecta. We therefore examine the effect of the inward diffusion of hydrogen and its effect on the molecular yield of the SNe. 

The molecular budgets of SN ejecta are determined by their physical conditions and composition. In \S2 we describe the relevant parameters: explosion energy, initial density, the ejecta mass, and composition of the supernovae under study. These are used to follow the evolution of ejecta temperature and density as a function of time.  The reactions taking place in the different layers of the ejecta depend on the degree of elemental mixing, and the section also describes the prescriptions we used to characterize this process. The different mechanisms pertaining to the ejecta chemistry are described in \S3. We first give a brief overview of the mathematical formalism, followed by a detailed description of the non-thermal destructive processes that operate in a SN environment. The results of our calculations for mixed and unmixed ejecta of various mass progenitors are presented in \S4, and in \S5 we briefly summarize and discuss the different implications of these results. Models for the formation of dust precursors in similar environments will be presented in a forthcoming paper (Cherchneff \& Dwek in preparation).

\section{MODEL INPUT PARAMETERS}

\subsection{Physical Conditions of the Ejecta}
Currently, there are no observational constraints on the Pop~III supernovae events and the evolution of their explosive ejcta. We therefore base our ejecta models on available theoretical explosion models, in particular those of Heger \& Woosley (2002), Umeda \& Nomoto (2002, hereafter, UN02) and NK03. Using these models, we derive simple analytical expressions for the gas parameters such as temperature, number density and velocity and consider various levels of mixing in the ejecta. Three progenitor masses are studied: two massive progenitors of mass 170 \Ms\ and 270 \Ms\ chosen as surrogates to PISNe, and one low mass 20 \Ms\ progenitor chosen to describe primordial CCSNe. 

The gas temperature $T$ in the post-explosion gas is determined mainly by the explosion energy which is released as kinetic energy into the gas. NK03 present various models of ejecta for CCSNe, PISNe and hypernovae. Each model is characterized by an explosion energy, a mass cut marking the division between the matter which remains in the core and that which is ejected and the \Ni\ mass produced in the explosion above the mass cut. By solving the radiative transfer equation taking into account the energy deposition by radioactive elements, NK03 derive temperature profiles for each of their massive progenitor ejecta. For our PISNe models, we chose the NK03 temperature profile of the oxygen-rich region for  their 170~\Ms~PISN unmixed case to describe our 170~\Ms~ejecta temperature variation with time. The temperature characterizing the helium core of PISNe modeled by Fryer et al. (2001) stays almost constant over the core extent. We then assume a constant temperature in the inner He core region so that temperature is independent of mass coordinate $M_r$ and simply fit the profile by the following power law of degree five
\begin{equation}
\label{temp1}
T( t)= T_0\times (A-B x+C x^2-D x^3 +E x^4-Fx^5),    
\end{equation}
where $T_0$ = 21,000~K is the gas temperature at some fiducial time $t_0$ = 100 days, $x = t/t_0$, and the fitting coefficients $A$, $B$, $C$, $D$, $E$ and $F$ are equal to 1.699, $8.568\times{10^{-1}}$, $1.761\times {10^{-1}}$, $1.762\times {10^{-2}}$, $8.229\times {10^{-4}}$, and $1.331\times{10^{-5}}$, respectively. Our 270~\Ms\ temperature profile is assumed to follow the same variation with time as given in eq. (\ref{temp1}). To account for the greater kinetic energy imparted by the explosion of a 270~\Ms\til progenitor, we multiply $T _0$ by a multiplying factor of 1.5. This value corresponds to the ratio of the central temperatures for the 170~\Ms\ and 270~\Ms\ progenitor models of  Heger \& Woosley (2002).  

For our CCSNe model, we assume that the gas temperature follows the variation of the C20 unmixed model of NK03. Assuming the ejecta follows quasi-adiabatic expansion, the temperature evolution with time is given by
\begin{equation}
\label{temp2}
T(t) = T_0\times \left({t\over t_0}\right)^{3(1-\gamma)}, 
\end{equation}
where $\gamma$ is an 'adiabatic' index. Using eq. (\ref{temp2}) to fit the NK03 C20 temperature profile gives a $\gamma$ value of 1.593. 

The ejecta expansion becomes homologous a few hours after explosion so that the gas density varies with time according to 
\begin{equation}
\label{rho}
\rho(M_r,\, t) = \rho(M_r,\,  t_0)\times \left({t \over{t_0}}\right)^{-3},
\end{equation}
where $M_r$ is the mass coordinate. As for temperature, and according to the PISN models of Fryer et al. (2001), we assume a constant gas density over the helium core so that no dependence with the mass coordinate is considered. The gas density profile chosen in the present study is that of NK03 for their 20~\Ms\ and 170        ~\Ms\ progenitors, assuming $\rho$(600~days) equals to $\rm 3\times 10^{-14}\til g\til cm^{-3}$ for all progenitor masses we study. 

The gas number density $n(t)$ for each model is given by:
\begin{equation}
\label{dens}
n(Mr,\, t) = \rho(M_r,\, t)/\mu_{gas}(t)
\end{equation}
where $\mu_{gas}(t)$ is the mean molecular weight of the gas, which varies with time as the chemical composition of the ejecta changes due to molecule formation.

For the sake of simplicity, we define a constant ejecta velocity $v$ determined by the explosion energy of the progenitor through its conversion to kinetic energy and given by
\begin{equation}
\label{vel}
 v = \sqrt{2\times E_0\over {M_{ej}}},
\end{equation}
where $E_0$ is the explosion energy and $M_{ej}$  is the mass ejected above mass cut during explosion. Explosion energies for the present models are listed in Table~1. The respective ejected masses are assumed to be equal to the progenitor masses as Pop.~III stars do not experience mass loss during their evolution due to the lack of dredged-up metals in their photospheric composition and the consequent wind acceleration through metallic lines. The velocity is assumed to be constant over the mass zones in the ejecta. 

We are interested in studying the chemistry from time $t_0$ = 100 days to t = 1000 days. The initial time range is justified by the appearance of CO, SiO and dust as early as 110, 160, and 450 days, respectively, in the ejecta of \objectname{SN1987A} (Catchpole \& Glass 1987, Danziger et al. 1991, Wooden et al. 1993). Our PISN ejecta are hotter than that of \objectname{SN1987A} but similar temperature regimes are encountered at times t $\ge$ 300 days. The final time is determined by the time when the gas density and temperature in the ejecta are too low to foster efficient gas-phase molecular formation. The ejecta parameters are summarized in Table~1 whereas the variation of the gas density and temperature with time t is illustrated in Figure 1 for the various progenitor masses considered. 

\subsection{Mixing in the Ejecta}
Mixing is likely to occur during explosion for the instability of the nickel bubble resulting in the development of Rayleigh-Taylor instabilities (Woosley 1988, Arnett 1988, Herant \& Benz 1991, M{\"u}ller et al. 1991, Kifonidis et al. 2003) and instabilities developing in the post-shocked regions of the propagating blast wave (Chevalier 1976, Bandiera 1984, Benz \& Thielemann 1990). Evidence for strong \Ni~mixing in the shell of \objectname{SN1987A} is found through observation of hard X-rays stemming from \Ni/\Co~decay and $\gamma$-rays down-scattering as early as 140 days after explosion (Itoh et al. 1987, Pinto \& Woosley, 1988, Sunyaev et al. 1990). Another direct evidence for heavy element mixing comes from the extraction and study of isotopically anomalous inclusions in meteorites. The presence of type X silicon carbide (SiC) grains and silicon nitride (S$_3$N$_4$) inclusions bearing the $^{44}$Ti supernova signature in meteorites suggests deep and inhomogeneous mixing between the various heavy element mass shells (Zinner 2006). As to light elements, hydrogen deep mixing down to the inner mass zones is invoked to explain the plateau shape of \objectname{SN1987A} light curve at  times greater than 80 days (Woosley 1988, Arnett \& Fu 1989, Shigeyama \& Nomoto 1990). However, H mixing is likely to occur at a macroscopic level with the formation of H-rich bubbles in the unstable layer located between the He-CO  and the H-He interfaces as early as a few hours after explosion (Fryxell et al. 1991, Herant \& Benz 1991, M{\"u}ller et al. 1991, Burrows \& van Riper 1995, Kifonidis et al. 2003). Some hydrogen diffusion may also occur at the base of the hydrogen envelope where helium and heavy element-rich fingers are simultaneously present (M{\"u}ller et al. 1991, Kifonidis et al. 2003). To circumvent the problem of the complexity of mixing, NK03 considered explosive models with and without mixing, which result in two extreme cases for their ejecta: a fully mixed gas and a stratified ejecta in which each layer reflects the prior nucleosynthesis stages of the progenitor, except for the inner most layer which is the result of explosive nucleosynthesis. We follow the same strategy in the present paper: for the two very massive progenitors, 170 \Ms~and 270 \Ms~and the core-collapse SN progenitor, 20 \Ms, we consider fully mixed ejecta while stratified ejecta are studied for the 170 \Ms~and  20 \Ms~progenitors.  

\subsection{Initial Ejecta Composition} 

The initial chemical compositions for the primordial supernova ejecta models are taken from fully mixed and unmixed explosion models. For fully mixed ejecta, mass yields are from UN02 and Heger \& Woosley (2002) for our massive 170~\Ms\ and 270~\Ms\ progenitors and from UN02 for the 20~\Ms\ CCSN model. The unmixed ejecta compositions for the 170~\Ms\ progenitor and the 20~\Ms\ progenitor are those of NK03.  We subdivide their helium cores in zones of distinct chemical composition. For the 170~\Ms\ progenitor, we consider five distinct zones: (1) is Si/S/Fe-rich  from 0 to 20~\Ms, (2) is O/Si/S-rich from 20 to 40~\Ms, (3) is O/Mg/Si-rich from 40 to 55~\Ms, (4) is O/C/Mg-rich from 55 to 78~\Ms, and finally (5) is O/C/He-rich from 78 to 82~\Ms. For the 20~\Ms\ progenitor, the zoning comprises four zones as follows: (1) is Si/S/Fe-rich and extends from 2.4 to 3~\Ms, (2) is O/Si/S-rich from 3 to 3.6~\Ms, (3) is O/C/Mg-rich  from 3.6 to 4.95~\Ms, and finally (4) is He/C/O-rich from 4.95 to 5.85~\Ms. For the sake of simplicity, we ignore the variations of the elemental abundances with mass coordinates and take them to be constant within each zone. 

We use these elemental mass yields to calculate the total number of elemental species, the gas initial molecular weight $\mu_{gas}(t_0)$ and gas number density  $n(M_r, t_0)$. We then derive the number density at 100 days post-explosion for each element, and use these data as initial conditions when solving our set of non-linear, coupled, differential equations (see section 3). These data along with the resulting initial gas mean molecular weight are tabulated in Table 4 for fully-mixed ejecta whilst Table 5 summarizes the post-explosion chemical composition as a function of zoning for our unmixed ejecta models. We then explore the effect  of hydrogen mixing by diffusion from the progenitor envelope for the 170 \Ms\ fully-mixed ejecta. In doing so, we set the H content as a free parameter and assume that H can microscopically mix within the heavy element-rich He core. Values of H mixing ranges from 0\% to 10\% of the total hydrogen envelope mass given by UN02.  

\section{EJECTA CHEMISTRY}

A chemical kinetic description of the ejecta is based on a gas initial chemical composition and a set of chemical reactions describing the chemical processes at play and applied to the ejecta physical conditions. For the high gas temperatures and densities characterizing our modeled ejecta, they include: (1) termolecular reactions efficient in high density media and where formation of molecules occurs through collision with the ambient gas which carries away the excess energy of the reaction. We also consider their reverse processes which are thermal fragmentation through collisions with the ambient gas at high temperatures; (2) bi-molecular processes, predominantly neutral-neutral (hereafter NN) reactions. The reactions with activation energy barriers require high temperatures to overcome the energy barrier whereas reactions without activation energy can proceed at lower temperatures; (3) ion-molecule reactions (formation/destruction reactions and charge exchange processes), which have no energy barrier and can therefore contribute at low gas temperatures; (4) temperature-independent radiative association reactions (hereafter RA) in which the formation of a species occurs through the emission of a photon which carries off the excess energy released in the formation of the adduct. 

Because of the unique nature of SN ejecta, being powered by the decay of radioactive elements and its high temperatures, we also consider the effects of various non-thermal processes including the destruction of molecules by the cascade of energetic electrons and UV photons that are generated by the down-scattering of  $\gamma$ rays in the ejecta. These non-thermal processes will be discussed in more details in the following section. 

The temporal variation of the number density of a molecular specie $i$ located in a given mass zone $M_r$ is described by the following rate equation:
\begin{equation}
\label{dndt}
{\partial n_i(M_r\, t) \over{\partial t}} = P_i -  L_i = \sum_j k_{ji} n_j n_i - \sum_k k_{ik} n_i n_k
\end{equation}

\noindent
where $P{\rm _i}$ are the production ($\equiv$  formation) processes and $L{\rm _i}$ are the loss ($\equiv$ destruction) processes for species $i$, and ${k_{ij}}$ and ${k_{ik}}$ are the temperature-dependent rates for reactions between species $i$-$j$ and $i$-$k$, respectively. The reaction rates $k_{ji}$ and $k_{ik}$ are written under the form of Arrhenius expressions such as 
\begin{equation}
\label{kij}
k_{ij}(T) = A_{ij} \times \left({T \over 300}\right)^{\nu}  \times \exp (-E_{ij} / T), 
\end{equation}

\noindent
where $T$ is the temperature given by equations (\ref{temp1}) and (\ref{temp2}), $A_{ij}$ the Arrhenius coefficient in s$^{-1}$ molecule$^{-1}$, cm$^3$ or cm$^6$ s$^{-1}$ molecule$^{-1}$ for a unimolecular, bimolecular or termolecular processes respectively, $\nu$ reflects the temperature dependance of the reaction, and  $E_{ij}$ is the activation energy barrier in K$^{-1}$. The ensemble of equations as in eq.(\ref{dndt}) represents a set of $N$ non-linear, coupled, ordinary differential equations to solve, where $N$ is the number of species (atoms, ions, molecules, and electrons) included in the chemical description of the ejecta. In total, the system comprises up to 79 species listed in Table~2 and between 400 to 500 reactions, depending on the ejecta region under study. The full chemical scheme is listed in Table~9. Reaction rates have either been measured under laboratory conditions or theoretically calculated using transition state theory. When not documented, the rates are estimated using 'educated' guesses. The NIST chemical kinetics database is used as primary source for NN processes, completed by the UDFA06 database (Woodall et al. 2007) when necessary. The numbering of some reactions specified in the text refers to that of Table 9. For dust precursor formation, processes and reaction rates stem from combustion chemistry studies, environmental and material sciences, and full details are given in Cherchneff \& Dwek  (in preparation). 

\subsection{Non-thermal Processes} 

\subsubsection{Destruction by Compton Electrons}
A supernova explosion produces $^{56}$Ni which rapidly decays to $^{56}$Co with a half life of 6~days, which in turn decays to $^{56}$Fe with a half-life of $\sim 77$ days. The decay of $^{56}$Co deposits 3.57~MeV in the form of \grays\ in the ejecta, which  powers the SN light curve.  (Woosley et al. 1989). Compton scattering degrades the \grays, which have an average energy of 1.24~MeV, to hard X-rays which through a cascade of inverse Compton, ionization and recombination processes degrade further into UV photons. The fast Compton electrons thermalize by heating, exciting and ionizing the ejecta, adding to the reservoir of UV photons. In \objectname{SN1987A} , the light curve between days 100 and 1000 could be reproduced if 0.075 \Ms~of $^{56}$Co was ejected in the explosion.

The fast Compton electrons and UV radiation can have a significant effect on the chemistry of the ejecta. Compton electrons were proposed in several studies to be one of the dominant destruction routes to molecules in SN ejecta (Liu \& Dalgarno 1994, 1995, 1996). Clayton et al. (1999, 2001) furthermore suggested that atomic carbon and carbon dust can be produced in oxygen-rich part of the ejecta by CO dissociation due to collision with Compton electrons. We therefore pay special attention to deriving destruction rates for similar radioactivity-induced processes in the ejecta of our primordial SNe. In addition, we will also explore the role of UV photons on the chemistry of the ejecta. 

Not all the radioactive decay is deposited in the ejecta. The early emergence of \grays~and X-rays a few hundred days after the explosion of \objectname{SN1987A} provides strong evidence that the ejecta in macroscopically mixed and presumably clumpy (McCray 1993). However, the fraction of this escaping energy is small, and throughout this paper we will assume that all the radioactive energy is deposited uniformly in the ejecta. The rate of energy deposition by thermalized \Co~\grays~in the ejecta of SN1987A is given by Woosley et al. (1989):
\begin{equation}
\label{lgamma}
L_{\gamma} = 9.54\times 10^{41} \times  \exp(-t/\tau_{56}) \times (1-\exp[-\tau_0\, (t/t_0)^{-2}]),
\end{equation}
where $L_{\gamma}$ is given in erg s$^{-1}$, $\tau_{56} = 111.26$~d is the e-folding time of $^{56}$Co decay, and $\tau_0$ is effective $\gamma-$ray optical depth of the ejecta at some fiducial time $t_0$.  $L_{\gamma}$ scales linearly with the mass of $^{56}$Co in the ejecta. Therefore, the destruction rate by Compton electrons for species $i$ in s$^{-1}$ per particle, $k_C$, can be written in terms of $M_{56}$(1987A), the mass of \Co~produced in \objectname{SN1987A}, as (Liu \& Dalgarno 1995, CL08):

\begin{equation}
\label{kc}
k_C(i)= {5.95\times 10^{53} \over W_i \times N_i}\, \left[ {M_{56} \over M_{56}({\rm SN1987A)}} \right] \times \exp(-t/\tau_{56}) \times (1-\exp[-\tau_0(t/t_0)^{-2}]),
\end{equation}

\noindent
where $N_i$ is the total number of particles of species $i$, $M_{56}$ is the mass of \Co\ in the ejecta of the SN being studied, and $W_i$ is the mean energy (in eV) per ion-pair, dissociation or excitation for species $i$. $W_i$ is defined as the ratio of the energy of the incident electron divided by the number of ionization, dissociation or excitation produced by collision with the incident electron until it comes to rest (Liu \& Dalgarno 1994, Dalgarno et al. 1999). 

The effective $\gamma-$ray optical depth $\tau(t)$ at time $t$ can be written as:
\begin{eqnarray}
\label{tau}
\tau(t) & \equiv & \kappa_{56}\times \phi(t) \\ \nonumber
 & = & \kappa_{56}\times \rho(t)\, R(t) = \kappa_{56}\times \left({3\, M_{He} \over 4\, \pi\, R(t)^2}\right) 
 \end{eqnarray}                                                                      
where $\kappa_{56}$ is the average $\gamma$-ray mass absorption coefficient in cm$^2$~g$^{-1}$, $\phi(t)$ is the mass-column density of the ejecta in g~cm$^{-2}$, $\rho(t)$ and $R(t)$ are, respectively, is the mass density and radius of the ejecta at time $t$, and $M_{He}$ is the mass of the helium core. The average mass absorption coefficient depends on the ejecta composition and the distribution of the radioactive material within the ejecta. For slabs of material consisting of pure He, C, O, Mg, Si, or Fe, $\kappa(E_{\gamma}=1.25$~MeV)$\approx 0.056$~cm$^2$~g$^{-1}$. For the mixed distribution of $^{56}$Co in model 10HMM of SN1987A, Woosley et al. (1989) derive an effective value of $\kappa_{56}=0.033$~cm$^2$~g$^{-1}$, which is the value that we will adopt for all models in this paper.
Table~1 lists the value of $\tau_0 \equiv \tau(t_0)$, for $t_0 = 100$~d, and the relevant parameters used in its calculation,  for the different primordial SNe used in this study.

The interaction of the Compton electrons with the molecules leads to their dissociation, ionization and fragmentation into ionic products. The branching ratios for the different processes depends on $W_i$, the mean energies per ion-pair for a given species. Available values of $W_i$ for molecules that can form in the ejecta are listed in Table 3. When data are not available, we just assume values similar to those for CO. Using the \Co\ mass listed in Table~1 for each progenitor, we calculate the time dependent rates from eq. (\ref{kc}).  However, the rate values need to be expressed under a Arrhenius temperature-dependent form as given by eq. (\ref{kij}). We thus compose the reverse of the time-dependent temperature functions given by equations (\ref{temp1}) and (\ref{temp2}) for our modeled ejecta with our rate function given by eq. (\ref{kc}), and derive Compton electron destruction rates as a function of ejecta temperature. We then fit those rates with a Arrhenius function as given by eq. (\ref{kij}). The corresponding Arrhenius parameters are listed in Table~3. 

\subsubsection{Destruction by UV Radiation} 

As they slow down in the ejecta, the high-energy Compton electrons deposit their energy in three channels: heating, excitation, and ionization. The ionization of the ejecta by the primary electrons creates secondary fast electrons, which also deposit their energy in these channels. The fraction of the energy going into each channel depends primarily on the ejecta composition, and on $x_e$, the electron fraction of the ejecta. Kozma \& Fransson (1992; hereafter KF92) calculated these quantities for different ejecta composition as a function of $x_e$ (Figs 3-5 in their paper). 
Of particular interest to our study is the fraction $\alpha$ of energy that is deposited in the ejecta that emerges as UV radiation. KF92 derived the value of $\alpha$ by calculating the evolution of $x_e$ for different layer compositions thus determining the fractional energy deposited by the high-energy electrons in the different channels as a function of time. They then calculated the amount of UV emission that is released by excitation and ionization in the different composition zones as a function of time. They found that the fraction of the energy that is deposited in the ejecta that emerges as UV photons with wavelengths below 3646~\AA\ is slowly rising from a value of $\sim 0.25$ on day 200 to $\sim 0.4$ on day 1000 (Fig 8 in their paper). The destruction rate of molecules by UV photons (in s$^{-1}$ molecule$^{-1}$) is thus given by:
\begin{equation}
\label{kuv}
k_{UV}(i)= {\alpha \times 5.95\times 10^{53} \over {E_{UV} \times N_i}}\, \left[ {M_{56} \over {M_{56}({\rm SN1987A})}} \right] \times \exp(-t/\tau_{56}) \times (1-\exp[-\tau_0(t/t_0)^{-2}]),
\end{equation}
where $\alpha$ is the fraction of deposited energy re-emitted as UV photons, and $E_{UV}$ is the energy of the UV photons. According to FK92, non-thermal excitations and recombinations in the oxygen-rich zone result in emitting OI 1302 \AA~and 1356 \AA~photons. We therefore calculate $E_{UV}$ as being the energy of a fiducial photon of wavelength 1302 \AA. Other terms are as in equation (\ref{kc}). 

\section{RESULTS}
 In the following sections, we present the abundances and the mass ejected at day 1000 of molecules produced in fully-mixed and unmixed ejecta of supernovae, excluding those that are dust molecular precursors. Results for gas-phase dust precursors and small clusters will be presented in a forthcoming paper (Cherchneff \& Dwek in preparation).  
 
 \subsection{Ejecta Chemistry is not at Steady State}

Previous studies of molecular formation in SN1987A assumed that the chemistry was at steady state, implying that species number densities did not vary with time (Lepp et al. 1990, Liu \& Dalgarno 1994, 1995, Gearhart et al. 1999, Clayton et al. 2001). Such assumption was also made by Todini \& Ferrrara (2001) and SFS04 in their studies of dust formation in primeval supernovae. When a small number of chemical reactions is considered, the steady state approximation may be valid, allowing for the direct derivation of analytical solutions for chemical abundances. This assumption requires that at early times, the rates of the chemical reactions be fast compared to the rate of density and temperature changes in the ejecta, so that chemical abundances would quickly reach their equilibrium value. However, fast chemistry does not 'a priori' ensure the validity of the steady state approximation in larger chemical systems. Indeed, such large systems bring a greater complexity in terms of chemical rates. Certain species may reach steady state abundances while others are still being formed or destroyed under non-equilibrium conditions.

Figure~2 illustrates this point for the 170~\Ms~ progenitor fully-mixed ejecta, by comparing select reaction rates to $k_{dyn}$, the inverse dynamical timescale, which is defined as:
\begin{equation}
 k_{dyn} = {1 \over {t}} = {v  \over {R(t)},}
\end{equation}
where $v$ is the ejecta velocity given by eq. (\ref{vel}) and listed in Table~1, and $R(t)$ is the ejecta position at a given time $t$ after explosion. The rates depicted in the figure correspond to those of the major chemical processes involved in the formation and destruction of carbon monoxide, CO. The formation processes are the radiative association reaction (RA4)
\begin{equation}
\label{co}
{\rm C + O \rightarrow CO + h\nu,}
\end{equation}
and the bimolecular process (NN56)
\begin{equation}
\label{co2}
{\rm C + O_2 \rightarrow CO + O.}
\end{equation}
The major CO destruction processes are the bimolecular process (NN70)
\begin{equation}
\label{sico}
{\rm Si + CO \rightarrow SiO + C,}
\end{equation}
the reaction with helium ions (IM13)
\begin{equation}
\label{heco}
{\rm He^+ + CO \rightarrow C^+ + O + He,}
\end{equation}
and the collision with $\gamma$-rays-induced Compton electrons (CED31) given by 
\begin{equation}
\label{coe}
{\rm CO + e^-_{Compton} \rightarrow C + O + e^-_{Compton}}. 
\end{equation}

We see from Figure 2 that prior to $\sim$ 480 days post-explosion, all reaction rates are larger than the dynamical rate ${\rm k_{dyn}}$. A fast chemistry takes place but reactions proceed more or less efficiently, resulting in a non-equilibrium chemistry and a drive towards formation of molecules. At times greater than 480 days, the RA rate becomes smaller than ${\rm k_{dyn}}$ and the reaction 'freezes' out  when bimolecular processes are still active in building up molecules. Compton electron and He$^+$ reactions are always important in destroying CO up to 900 days and 700 days post-explosion, respectively. Therefore, the relevant processes to CO chemistry are activated and switched off at different times in the ejecta, and the chemistry is by no means at steady-state over the time period of interest. Consequently, molecular abundances show strong variations with time regardless of the initial composition and mixing in the ejecta. This point will be discussed in more detail below.

\subsection{Fully-mixed Ejecta}

We first consider the chemistry of the fully-microscopically-mixed helium cores for the primordial SN surrogates under study, exploring the impact of hydrogen mixing on the ejecta chemistry and the dependence of results on the progenitor mass. No UV destruction rates have been included in these models. 

\subsubsection{Impact of Hydrogen Mixing} 
Molecular abundances for the 170 \Ms\til surrogate are shown in Figure 3 considering two cases: (a) when no hydrogen is mixed into the He core, and (b) when 10\% of the hydrogen mass of the progenitor's envelope is microscopically-mixed uniformly throughout the He core. In the absence of hydrogen, there exist two phases of molecular formation as illustrated in Figure 3a. The first phase arises at $\sim$ 350 days when three molecules dominate the ejecta: silicon monoxide, SiO, silicon sulfide, SiS, and carbon monoxide, CO. For SiS and CO, the dominant building process is radiative association accounting for more than 94 \% of the total formation rate, whereas for SiO, RA accounts for only $\sim$ 56\% of the formation rate. The extra formation pathway is the Si reaction with CO  leading to the buildup of SiO  from CO. Destruction of SiS and SiO occurs through reactions with He$^+$ at the same time that SiO is being formed by the destruction of CO.  At times close to 440 days, silicon-based dust precursors (i.e., small (SiO)$_n$ and (SiO$_2$)$_n$ clusters) start forming due to the cooler gas  temperatures, and thus deplete Si-bearing molecules and SiO. Reaction with He$^+$ remains an important destruction channel for SiS, SiO and CO but for the latter, fast Compton electrons also contribute to the destruction. The second phase of molecular formation starts at $\sim$ 700 days when most of SiO is depleted into dust precursors.  CO molecules keep forming from RA and NN reactions involving atomic oxygen and carbon. The formation of O-bearing species like O$_2$ and SO are mainly triggered by NN processes involving atomic oxygen and sulphur. Molecular destruction is driven by NN processes as well since He$^+$ has recombined to neutral and the level of ionization has decreased by a factor of 100 in the ejecta. To summarize, the first phase of molecular formation at high temperatures is triggered by RA processes involving atomic species, whereas the second low-temperature phase is driven by NN reactions with small activation energy barriers. 
 
When hydrogen penetration from the progenitor envelope is allowed to a level of 10 \% of the progenitor hydrogen mass, an active and complex NN chemistry induced by radicals such as OH and CH, comes to play. The radical reservoir is fed by the products of reaction between heavy elements and molecular hydrogen. These radicals boost at early post-explosion times the formation of molecules like SO, NO, and O$_2$, the latter contributing to CO formation via reaction with atomic carbon. Although some of the processes mentioned in the previous section (i.e., reaction with He$^+$, RA processes) are still active, the whole kinetics is dominated by NN processes with and without activation energy barriers. Indeed, we see form Figure 3b that molecular formation is already fully developed at day 300, converting a large fraction of the ejecta into molecules. After 700 days, CO is converted to CO$_2$ from its reaction with OH and the ejecta is composed of O$_2$, SO, CO$_2$ and H$_2$. 
The molecular masses for the dominant species ejected at day 1000 are listed in Table 6 for the fully-mixed ejecta as well as the efficiency at forming molecules defined as the ratio of the molecular mass to the helium core mass. For the H-free fully-mixed ejecta of the 170 \Ms~progenitor, the dominant ejected species are CO and SO and the total molecular component of the ejecta equals $\sim$ 11.3 \Ms. Defining the molecule formation efficiency as the fraction of the ejected He-core mass that is converted to molecules, we get a formation efficiency of $\sim$ 13.7 \%. As expected, its H-rich counterpart produces a much larger molecular mass. It chiefly forms O$_2$, SO and CO$_2$ and the total molecular mass is $\sim 47$ \Ms, corresponding to a formation efficiency of $\sim$ 57 \%. 

Models assuming fully microscopically mixed ejecta are highly unrealistic, since observations of young SN remnants show that their ejecta are unmixed. Our purpose in admixing H is to illustrate the paramount effect of hydrogen on molecular synthesis in the ejecta. Hence, any hydrogen that microscopically mixes with nearby heavy elements at the interface of finger structures and inhomogeneities should induce efficient synthesis of H-bearing radicals and molecules.  Conversely, He$^+$ is detrimental to molecular formation, as first stressed by Lepp et al. (1990) in their attempt to modeling CO in SN1987A. Indeed, its attack represents one of the principal destruction channels to molecules whereas attack by Compton electron always plays  a minor role. This may not be the case for the unmixed ejecta, as discussed in the next section.

\subsubsection{Impact of Progenitor Mass}

We now turn to study the impact of the progenitor mass on the ejecta chemistry. We consider a massive progenitor surrogate of mass 270 \Ms\til so that we can directly compare results to those obtained in the previous section for the 170 \Ms\til PISN progenitor. We also consider a CCSN with a 20 \Ms\til progenitor to see if lower mass ejecta can efficiently form a molecular phase or not. As shown in Tables~1 and 4, different masses for Pop.~III stars imply different initial chemical compositions, explosion energies, ejection velocities, gas temperatures, and number densities. The gas column depths and opacities will change too, resulting in the Compton electron destruction rates listed in Table 4.  In both cases, no hydrogen penetration into the helium core is included. Molecular abundances with respect to total gas number density are presented in Figure 4. 

From the comparison of Figures~4 and 3 we see that the same species form in the 170 \Ms\til and the 270 \Ms\til ejecta. However, molecular formation is delayed for the more massive ejecta to day 550 primarily due to the higher gas temperatures. The chemical processes at play are identical to those mentioned in Section 4.2.1. SiO is too depleted on day 500 at the onset of dust precursor nucleation. As to molecular abundances, they are globally lower for the 270 \Ms\til progenitor due to a less favourable initial chemical composition of the ejecta gas at 100 days post-explosion. Indeed, we see from Table 4 that the mass of helium (and thus He$^+$) relative to the total progenitor mass is larger for the 270 \Ms\til surrogate compared to the 170 \Ms\til progenitor. The oxygen and carbon contents are also lower, resulting in smaller CO abundances in the ejecta. These conditions conspire to delay the formation of molecules, which consequently takes place at lower gas densities resulting in lower formation efficiencies. The results are summarized in Table 6 which compares the molecular yield of the 170 \Ms~and the 270 \Ms~PISNe without and with hydrogen mixing. Therefore, low-mass PISNe coming from Pop.~III progenitors should form more molecules in their ejecta than their very massive counterparts. 

It is of interest to compare the present results for the 170 \Ms\til and 270 \Ms\til surrogates to the study of dust formation in the fully-mixed ejecta of PISN by SFS04. Their modeling of dust formation accounts for the formation of CO and SiO following the formalism of Todini \& Ferrara (2001). However, both studies derive CO and SiO masses assuming those molecules form at steady state, and consider only  one formation channel (RA reaction) and one destruction channel via collision with Compton electrons. SFS04 derive a total CO mass of $\sim 6$ and 0.01 \Ms\ in the ejecta of the 170~\Ms\ and 260~\Ms\ PISN, respectively, but they do not show the evolution of molecular mass with time. In our model, the CO mass ejected at 1000 days after the explosion of the 170~\Ms\ progenitor is 5.8~\Ms, seemingly in good agreement with the SFS04 value. However, we ascribe this result to pure coincidence, since the chemical approaches and physical models used in our studies are totally different from theirs. In particular, SFS04 assume much lower temperatures in their PISN ejecta than ours. Furthermore, their initial carbon yield is that of Heger \& Woosley (2002) which is twice as large as our initial carbon mass taken from Umeda \& Nomoto (2002), suggesting a lower CO formation efficiency in their model. There is also a great discrepancy between the CO yield calculated by SFS04 for their 260~\Ms\ progenitor, and our CO yield for the 270 \Ms\ progenitor, which is about 300 times larger. We see from Figure 4 that CO formation proceeds at times greater than 500 days via non-steady state chemistry. NN bimolecular processes such as the reaction of atomic carbon with molecular oxygen commands CO formation whereas the dominant destruction channel is He$^+$ attack. These mechanisms are not considered in the SFS04 study. All shortcomings in the SFS04 models point out the importance of using a kinetic approach to {\it both} processes, the synthesis of molecules and the nucleation of dust in the ejecta. The two are intrinsically linked, since the formation of molecules (that are not dust precursors) depletes the ejecta from refractory elements that would otherwise be included in the dust formation process. 

In the 20 \Ms\til case, molecular formation patterns are quite different. On one hand, the lower ejecta temperatures foster molecular synthesis. On the other hand, the large helium content of the ejecta and the resulting higher  He$^+$ abundances efficiently inhibit molecule production. The initial elemental composition also implies less refractory elements like Si or Mg available for the formation of molecules and the nucleation of metal oxides. The combination of the paucity of refractory elements and the elevated He$^+$ abundances at times less than 700 days explain the low molecular abundances, as illustrated in Figure 4. Once He$^+$ has recombined at late times, the formation of molecules like CO, SO and O$_2$ can proceed. We see from Table~6 that the fully-mixed ejecta of CCSNe are as efficient as their massive counterparts at synthesizing molecules at late post-explosion times.

\subsection{Unmixed Ejecta}

As stressed in the previous Section, it is unlikely that SN ejecta are fully microscopically-mixed. We thus consider totally unmixed ejecta in which each mass zone of the helium core has retained the stratified pre-explosive stellar composition, except for the inner most mass zone whose composition reflects the explosion nucleosynthesis. We further assume that the elements are microscopically-mixed within each mass zone. Results for the unmixed ejecta of two 'primordial' SNe, one 170 \Ms~PISN and a 20 \Ms~CCSN, are presented in the following sections. The zoning and initial elemental compositions are those of Table 5, and we explore the impact on the chemistry of a secondary ultraviolet (UV) field as defined in Section 3.1.2. 

\subsubsection{170~\Ms\ Ejecta: Impact of UV Radiation}

Molecular abundances normalized to the total gas number density are shown in Figure~5 for zones~1 and 2, in Figure~6 for zones~3 and 4, and in Figure~7 for zone 5. In zone~1, the formation of S$_2$ is coupled to that of  SiS and FeS. The major formation process for SiS is the NN72 reaction S$_2$ + Si $\longrightarrow$ SiS + S , whilst the RA reaction between atomic S and Si contributes to a lesser extent. The reverse reaction of the NN channel is actually the major formation process for S$_2$. However, since the initial mass of Si is four times larger than that of S, the net formation of SiS is always more efficient than that of S$_2$. After $\sim 350$~days, the reservoir of S$_2$ is depleted through the continuous formation of SiS and the simultaneous formation of FeS, the molecular precursor to iron sulfide clusters, through the reaction of Fe  with S$_2$ (see Figure 5a). As a result, SiS is the most abundant molecules formed in zone 1.

In zone~2, most molecules are rapidly formed from RA processes whereas thermal fragmentation is the dominant destruction process at early times, due to the high gas temperatures. O$_2$ and SiO formation are coupled, for O$_2$ is partly destroyed by the NN69 reaction with atomic Si to form SiO. At $\sim$ 250 days post-explosion, SiO abundance decreases due to nucleation of silica dust precursors such as (SiO)$_n$, resulting in a sharp increase of O$_2$ abundance. As SO and CO are also formed in the NN reaction of O$_2$ with S and C atoms (reactions NN76 and NN56, respectively), their abundances show a steep rise with increasing O$_2$. The low amount of CO formed in the gas is due to the initial low carbon content of this mass zone. 

The chemical processes pertaining to zone~2 also apply to zone~3 and the resulting molecules are similar to those of zone~2, as seen in Figure~6a. The differences in abundance variation and magnitude with time are due to the different initial chemical composition of zone~3. Specifically, the higher carbon initial mass yield drives the formation of CO via RA process at early times, depleting some of the atomic oxygen available to the formation of O$_2$ and resulting in higher CO abundances and lower O$_2$ and SO abundances over time than in zone~2.

Despite the fact that zone~4 is oxygen-rich, it is characterized by high initial carbon mass compared to zones~2 and 3 (see Table 5). However, it is expected that its initial chemical composition precludes formation of carbon-rich molecules other than CO as its C/O ratio is $\sim$ 0.29. However, we see from inspection of Figure 6b that very early on, C$_2$ and C$_3$ form in large amounts. This carbon chain formation is a consequence of the thermal fragmentation of CO at very high temperatures and its rapid conversion to C$_2$. Indeed, fragmentation studies of CO in shock tube experiments at temperatures between 5000 K and 18000 K show that C$_2$ and electronically-excited CO$^*$ are always detected along with the collisional destruction of CO in the high temperature post-shock gas (Fairbairn et al. 1968, Appleton et al. 1970, Hanson 1973). This is readily explained by the following chemical mechanism (labelled TF13, NN57 and TF18 in Table 9):
\begin{equation}
\label{com}
{CO + M \leftrightarrow C + O + M,}
\end{equation}
\begin{equation}
\label{cco}
{C + CO \leftrightarrow  C_2 + O,}
\end{equation}
and 
\begin{equation}
\label{c2m} 
{C_2 + M \leftrightarrow C + C + M.}
\end{equation}
where M is the colliding buffer gas (chosen as Ar, O, C, or CO in shock tube experiments), and all reactions and their reverse processes occur simultaneously. We see that along with the destruction of CO by collision with M, its rapid conversion to C$_2$ occurs through reaction with atomic carbon [eq. (\ref{cco})]. Chemical rates characterizing these reactions have high activation energy barriers and thus only proceed at high temperatures. Indeed, at temperatures less than 5000 K, rates for reactions (18), (19) and (20) become small. As to C$_3$, the build-up of the end-product of the carbon chains is due to its formation from reaction of two C$_2$ molecules which rapidly depletes C$_2$ chains from the gas, as seen in Figure 6b. A carbon dust mass yield can not be derived from the C$_3$ abundance profile, for the formation of C-bearing large chains and rings has not yet been included in the chemical network. It is expected that C$_3$ will be transformed in larger carbon chains and that atomic oxygen in the gas will partly destroy those chains through the formation of CO. Therefore, carbon chain abundances should be smaller than values for C$_3$ derived in this study.  However, the quick conversion of CO into C$_2$ will keep triggering the build-up of carbon dust molecular precursors in this O-rich mass zone. Two rates are reported in the literature for the C~+~CO reaction: a fast rate calculated by Fairbairn et al. (1968) and Hanson et al. (1973), which we use in our ejecta model, and a value listed in the UDFA06 database but with no traceable reference (Woodhall et al. 2007). The latter is $\sim$ 10 times smaller than the former. We ran the full model for zone~4 including the lowest rate and find that the C$_3$ abundance at day 1000 is decreased by a factor of 3, the CO abundance is increased by a factor of 6, and the O$_2$ abundance is decreased by 30\% compared to the values illustrated in Figure 6b. In both cases however, the quick CO conversation to C$_2$ via the C~+~CO reaction observed in thermal fragmentation experiments is a 'natural' mechanism to produce carbon-bearing species and carbon dust in a hot, oxygen-rich environment where carbon is initially in atomic form. When CO conversion to C$_2$ is not considered, no C$_2$ forms in large enough quantities to provide significant amounts of the end-product carbon chain C$_3$. 

To account for the formation of carbon dust in CCSNe, Clayton et al. (1999, 2001) proposed that CO dissociates in a fully-mixed ejecta from collisions with Compton electrons and reactions with He$^+$. The chemistry is assumed to be at steady state and CO dissociation creates a pool of carbon atoms from which carbon chains and solid clusters can nucleate and condense. However, we see from Section 4.1 that under non-steady state conditions, the two mechanisms invoked by Clayton et al. do not succeed in destroying enough CO to provide free carbon atoms. Indeed, for the three hydrogen-free mixed ejecta, CO is always the dominant species synthesized in the gas. Furthermore, He$^+$ being a predator to molecules at early times, CO formation is delayed to post-explosion times characterized by temperatures low enough to preclude the quick CO conversion to C$_2$ from operating. We conclude that fully-mixed  SN ejecta can not form free atomic carbon and carbon-based solids in general. Conversely, unmixed SN ejecta, in particular He-free zones where oxygen is more abundant than carbon, can generate carbon chains from the quick conversion of CO through the C~+~CO reaction operating at high temperatures. As to prevalent chemical formation processes in zone~4, we find that both RA reactions and NN bimolecular routes are active at high temperatures to forming species which are destroyed by thermal fragmentation reactions. Molecular formation and destruction in the intermediate temperature regime (1000 K - 5000 K) are too governed by NN bimolecular reactions when destruction via collisions with Compton electron also participate but to a much less extent. 

As to zone 5, we see from Figure 7 that the major molecule to form is carbon monoxide. In the outer zone, the formation of He$^+$ at early times triggers the simultaneous formation of ions like O$^+$ and C$^+$. NN processes are then not as important as they are in other zones for the formation of molecules. This fact, combined with the destruction of species by He$^+$, hampers the effective formation of molecules over most of the post-explosion times. CO, and to a less extent O$_2$, thus form at late times after He$^+$ recombination.

The molecular masses ejected at day 1000 are summarized for all zones in Table 7 and illustrated in Figure 8. The efficiencies listed are defined as the ratio of the molecular mass formed per zone to the zone mass. The value in the last column corresponds to the total molecular mass formed in the unmixed ejecta divided by the summed mass of the zones. In the unmixed case, molecules reflect the chemical composition of the zone in which they form. Dioxygen O$_2$ is by far the most abundant species in the ejecta  and form efficiently in three mass zones when SiS too forms in large amount but is confined to the innermost region of the He core. The final molecular content of the ejecta equals $\sim$ 37.1 \Ms, and the total formation efficiency is 45.07 \%, a very high value. It means that massive PISNe are effective at converting almost half of their initial atomic content into a molecular phase. This value is much larger than that for the fully-microscopically mixed case without hydrogen mixing. This result once again illustrates the primary role of He$^+$ in impeding the formation of molecules. 

The impact of the UV field as described in Section 3.1.2 is explored for zone 4 which produces the largest amount of molecules. The results on molecule abundances are shown in Figure 9. Formation and destruction processes for the dominant species are similar to those when no UV field is considered. While thermal fragmentation and NN bimolecular reactions remain the dominant destruction processes at high temperatures, molecular destruction by UV becomes effective at times greater than $\sim$ day 350, reducing the abundances of some important molecules. Specifically, O$_2$ is affected as its formation is postponed to later times, as seen in Figure 9. The final O$_2$ abundance at day 1000 is decreased by 30 \% compared to its value when no UV radiation is considered. However, the overall major chemical processes involved in molecular synthesis remain identical. We also consider UV photodissociation in our fully-mixed SN ejecta and find that the impact of UV photodissociation is minor compared to the destruction of chemical species by He$^+$. We thus conclude that UV photoprocesses if present have some minor impact as destruction channels to molecules and that their effect on the overall chemistry of the ejecta is limited. 

\subsubsection{20~\Ms\ Ejecta}
 We now turn to studying  the unmixed ejecta of a zero-metallicity, 20~\Ms~progenitor core-collapse supernova. Results for zones~1, 2 and 3 as defined in Table~5 are presented in Figures~10 and 11. These zones provide most of the molecules in the ejecta as seen from Table 8, which gives the total molecular masses formed in each zone at day 1000. These mass yields are illustrated in Figure 12.  
 
The chemical processes which occur in zone~1 are similar to those at play in zone 1 of the 170 \Ms~unmixed ejecta, resulting in the ejection of large amounts of SiS a day 1000. 
However, large amounts of oxygen and carbon compared to silicon and sulphur are initially presented in the initial composition of the 20 \Ms~CCSN, resulting in an active carbon chemistry. For example, CS forms in large quantities from the reaction of atomic C with S$_2$ and CO  abundances are too enhanced compared to zone 1 of the 170~\Ms~ejecta.
 
The chemistry of zone 2 is similar to that at play in zone 2 of the 170~\Ms~PISN, that is, early formation of most molecules by RA reactions and destruction by thermal fragmentation, and coupled chemical processes for the formation of O$_2$, SiO, CO and SO at later times. The dominant species ejected at day 1000 are O$_2$, SO and CO, as illustrated in Figure 10b.
 
Zone 3 can be compared to zone 4 of the 170~\Ms~progenitor case as both zones are characterized by a C/O ratio less than 1 (0.33 and 0.29 for the 20~\Ms~zone 3 and the 170~\Ms~zone 4, respectively) and are helium-free. Similar chemical processes are effective at building up molecules in both zones. The rapid conversion of CO to C$_2$ once again triggers the formation of carbon chains in zone 3 but the overall process is less efficient than for the 170~\Ms~progenitor owing to the lower temperatures in the ejecta. This is illustrated in Figure 10b where CO always remains more abundant than the carbon-chain end product C$_3$, and is gradually converted into O$_2$ at late post-explosion times by its reaction with atomic oxygen. 

In zone 4, characterized by a large He mass and a C/O ratio of 29.5, the mass of molecules formed is negligible as seen from Table 8. This is primarily due to the initial chemical composition of the zone where more than 98\% of the mass is helium while C and O only represent 1.7\% and 0.08\% of the zone mass, respectively. Molecules once formed are chiefly destroyed by He$^+$ at any times in the ejecta owing to its overwhelming presence. 

Table 8 and Figure 12 show that low-mass, zero-metallicity progenitors are almost as efficient at forming a molecular phase in their ejecta than their massive counterparts. The total efficiency at forming molecules is 35.65\%, a slightly lower value than that of the 170~\Ms~case. This lower efficiency is primarily due  to the 20~\Ms~zone~4, which, as discussed above, does not form molecules. 

\section{SUMMARY AND DISCUSSION}

We have investigated the chemistry of the ejecta of Pop.~III progenitor SNe and find that copious amounts of molecules form in these inhospitable environments. Of particular importance are the following points:
\begin{itemize}
\item As already stated by CL08, the chemistry in SN ejecta is not at steady state from 100 to 1000 days after explosion. New chemical channels involving neutral-neutral processes prevail over radiative association reactions, Compton electron destruction routes and photodissociation by ambient ultraviolet photons. Ion-molecule reactions play a role at late times when the ejecta is cool and diffuse. Our results for relevant molecules like CO or SiO disagree with existing studies due to the fact that the chemistry is not at steady state.  We find that the chemistry of Pop.~III SN ejecta, owing to the large ranges of temperatures and densities spanned over relatively short times, is complex, manifold, and conducive to molecule synthesis.

\item A new pathway to the formation of carbon chains is active in the oxygen-rich mass zone of the unmixed ejecta and is identified as the CO conversion to C$_2$ via collision with C. This fast conversion is usually observed in the thermal fragmentation of carbon monoxide in high temperature shock tube experiments. When this conversation is suppressed, no carbon-bearing molecules and  chains are formed. Thus, this conversion channel triggers the formation of carbon chains and dust in an oxygen-rich gas. It is then relevant to any gaseous O-rich environment characterized by high temperatures (T $>$ 5000 K)  and a large atomic carbon fraction.
  
\item The present results are extremely sensitive to mixing in the ejecta. We find that the injection of hydrogen from the progenitor envelope in fully-mixed ejecta boosts molecular synthesis via the formation of radicals like OH. This result too applies to unmixed ejecta. On the other hand, we show that helium severely hampers the formation of molecules through He$^+$ attack. Therefore, the detection of molecules in SN ejecta brings evidence for the non-mixing of helium with other elements in the ejecta gas. 

\item The results of our calculations show that the fully-mixed 170~\Ms~and 270~\Ms\ progenitors produce 11~\Ms~and 3.2~\Ms\ of molecules, respectively. Therefore, a larger progenitor mass does not imply a larger molecular content of the ejecta. This is chiefly due to the harsh physical conditions encountered in the ejecta of the 270~\Ms~progenitor and to its initial chemical composition. Indeed, although the 270 \Ms~model is characterized by larger heavy element masses with respect to its 170 \Ms~counterpart, its helium mass is almost three times greater than that of the 170~\Ms~case, implying efficient destruction of molecules. The admixing of 10\% of the hydrogen present in the 170~\Ms\ progenitor envelope into the fully-mixed ejecta dramatically increases its molecular yield to $\sim 47$~\Ms. The more realistic unmixed ejecta of a 170~\Ms\ progenitor supernova synthesizes $\sim 37$~\Ms\ of molecules at post-explosion day 1000, which is significant. About half of the initial elemental content of the He core is converted into molecules. The most abundant species by mass is O$_2$ followed by SiS, CO and SO. Its 20~\Ms\ counterpart produces  $\sim 1.2$~\Ms. O$_2$ is the dominant species followed by CO, SiS, and SO. This lower efficiency at forming chemical species for the low mass CCSN is due to the existence of its extended helium-rich outer zone in which molecular synthesis is suppressed. 

\item The present results hold for Pop.~III SNe but the large mass yields of molecules formed in their ejecta address the possibility of potential observational detection of new molecules in nearby SN ejecta.  As stated above, our primordial 20~\Ms\ progenitor forms O$_2$, CO, SiS, and SO shortly after explosion. CO and SiO have already  been detected at IR wavelengths in several SN ejecta. However, molecules like SiS and O$_2$ are tracers of stratified, unmixed ejecta while microscopic mixing with hydrogen produces tracer species like CO$_2$, OH and H$_2$O. Search for those chemical species should be undertaken at IR and submillimetre wavelengths. 
 \end{itemize}
 
The large amounts of molecules synthesized in the PISN ejecta are exposed to a harsh environment generated by the PISN blast wave. The shock expanding into the ambient circumstellar/interstellar medium generates fast particles that are accelerated to cosmic rays energies. These penetrate the cavity generated by the blast wave subjecting the ejecta to the constant bombardment by high energy electrons and nuclei. The pressure of the material that is shock-heated by the advancing blast wave will generate a reverse shock that will move into the expanding PISN ejecta (e.g.,~Nozawa et al. 2007). Locally, observations of young supernova, such as Cas~A, show that their ejecta is very clumpy, the clumps consisting of X-ray filaments with densities $\sim 1-10$~cm$^{-3}$,  and optical- and IR-line emitting knots with densities of $\sim 10^3-10^4$~cm$^{-3}$ (Fesen et al. 2006, Smith et al. 2009).  Therefore, the fate of the ejected molecules will depend on their environment. 
Low density filaments are heated by the reverse shock to X-ray emitting temperatures. Chemical species are not likely to survive this harsh environment. The dust and molecules inside the optical- and IR-line emitting clumps will encounter a much slower shock. Their fate will depend on the relative timescales of many different processes operating in the cavity of the young remnant: the radiative cooling time of the shocked clump, the timescales for the heating of the clump by thermal conduction and ambient cosmic rays, the evaporation timescale of the clump, and the timescale for the development of various instabilities that can lead to its fragmentation and subsequent evaporation. Any surviving molecules will be further subjected to the general diffuse interstellar UV radiation field generated by the  Pop~III stars. It is therefore very unlikely that the molecules synthesized in the Pop~III SN ejecta will have any global cosmological impact. However, they can have a significant local impact. The expanding PISN blast wave will generate during the radiative phase of its evolution a cold dense shell. This shell may be subject to various instabilities that can cause its collapse, forming the next generation of stellar objects (MacKey et al. 2003, Schneider et al. 2006). These stars, commonly referred to as Pop~2.5 stars, will form out of a gas that contains an admixture of the heavy elements, molecules, and dust that formed in the PISN ejecta. The mass of these stellar objects will depend on their ability to fragment into smaller structures, which is greatly facilitated by the cooling rate of the gas via atomic and molecular processes, and by the conversion of the cloud's internal energy to infrared emission by dust (Bromm \& Larson 2004). The survival of the molecules and dust, and their effect on the formation of stars in this propagating star formation scenario will be explored in a subsequent paper.

\acknowledgments IC would like to acknowledge support from the Swiss National Science Foundation through a Maria Heim-V{\"o}gtlin fellowship.

\clearpage

\begin{deluxetable}{lccc}
\tabletypesize{\scriptsize}
\tablecaption{Primordial supernova parameters used in this study (adapted from UN02, Heger \& Woosley 2002, and NK03.)}
 \tablewidth{0pt}
 \tablehead{
\colhead{  } & \colhead{20 \Ms} & \colhead{170 \Ms} & \colhead{270 \Ms}  
}
\startdata
E$_{\rm explosion} $(Ergs)& 1$\times 10^{51}$& 2$\times 10^{52}$& 8$ \times 10^{52} $\\
He core mass (\Ms) & 5.8 &82.3 & 129  \\
M($^{56}$Co) (\Ms)&  0.07 & 3.6& 9.8\\ 
v  (Km s$^{-1}$) & 2242 & 3439 & 5458  \\
T$_0$  (K) & 18000 & 21000& 31500\\ 
$\tau_0$ (g cm$^{-2}$)& 23.80 &146.07 & 90.89 \\
 \enddata
\end{deluxetable}
\clearpage


\begin{deluxetable}{lccccccccccc}
\tabletypesize{\scriptsize}
\rotate
\tablewidth{0pt}
\tablecaption{Species included in the ejecta chemical models}      
\tablehead{
\colhead{Atoms} & \colhead{H} & \colhead{He} & \colhead{O} &  \colhead{C}& \colhead{Si}& \colhead{S}& \colhead{Mg}& \colhead{Fe}& \colhead{Al} & \colhead{} & \colhead{}
}
\startdata  
 Diatomic  species& H$_2$ & OH & O$_2$ & CO & SiO & SO&NO  & MgO &FeO & AlO  &C$_2$  \\
 & CS&  CN & SiH  &SiC & Si$_2$ & SiS & SiN  &SH & N$_2$ & NH & MgS  \\
 & Fe$_2$  & & & & & & & & & & \\
\tableline
Tri-atomic species& H$_2$O & H$_2$S & HCN & CH$_2$ & C$_2$H &HCO &C$_3$ & CO$_2$ & OCS & OCN  & SiC$_2$  \\
 & Si$_3$&SiO$_2$&SO$_2$& NO$_2$&Fe$_3$& & & & & \\
\tableline
4-atom species &  Si$_2$O$_2$ & Mg$_2$O$_2$ & Mg$_2$S$_2$ &Fe$_2$O$_2$ & Fe$_2$S$_2$ &H$_2$CC &Fe$_4$&Si$_4$&C$_3$H  & CH$_3$  \\
\tableline
$\ge 5$-atom species& Si$_3$O$_3$ &Si$_2$O$_4$ &Si$_3$O$_6$&C$_3$H$_3$&C$_4$H$_4$& C$_6$H$_5$&C$_6$H$_6$ &  \\  
\tableline
Ions & H$^+$ & H$^-$ & He$^+$ & O$^+$&Si$^+$&S$^+$&Mg$^+$ &Fe$^+$&Al$^+$ & H$_2^+$  &H$_3^+$  \\
 & HeH$^+$& C$_2^+$& CO$^+$ &SiO$^+$ &SO$^+$ & H$_2$O$^+$ & HCO$^+$ & & &   \\                            
\enddata
\end{deluxetable}

\clearpage

%
\begin{deluxetable}{clccccc}
\tabletypesize{\scriptsize}
\tablecaption{Compton electron-induced reactions, corresponding mean energy per ion pair $W_{i}$ and Arrhenius coefficient A as a function of ejecta model.                                                                                                                                                                                                                                                                                                                                                                                                                                                                                                                                                                                                                                                                                                                                                                                                                                                                                                                                                                                                                                                                                                                                                                                                                                                                       }
 \tablewidth{0pt}
 \tablehead{
\colhead{ Species } &\colhead{Reaction} & \colhead{W$_{i}$ (eV)}  & \colhead{A - 20 \Ms}\tablenotemark{a} & \colhead{A - 170 \Ms}\tablenotemark{a} & \colhead{A - 270 \Ms}\tablenotemark{a} & Reference
}
\startdata
CO &$ \rightarrow$ O$^+$ + C& 768 &1.1610$ \times 10^{-7} $ & 9.4576$\times 10^{-7}$& 1.6741$\times 10^{-6} $& Liu \& Dalgarno (1995)\\
 & $ \rightarrow$ C$^+$ + O &247 & 3.6100$ \times 10^{-7} $&2.9406$ \times 10^{-6} $&5.2053$ \times 10^{-6} $& "  \\
 & $ \rightarrow$ C + O & 125 & 7.1333$ \times 10^{-7}$ &5.8107$ \times 10^{-6}$ & 1.0286$ \times 10^{-5}$ & " \\
 & $ \rightarrow$ CO$^+$ + e$^-$ & 34 & 2.6225$ \times 10^{-6}$ &2.1363$ \times 10^{-5}$ & 3.7815$ \times 10^{-5}$ & " \\
 \tableline
O & $ \rightarrow$ O$^+$ + e$^-$ & 46.2 & 1.9300$ \times 10^{-6}$ &1.5722$ \times 10^{-5}$ & 2.7829$ \times 10^{-5}$ & " \\
\tableline
C& $ \rightarrow$ C$^+$ + e$^-$ & 36.4 & 2.4496$ \times 10^{-6}$ &1.9954$ \times 10^{-5}$ & 3.5321$ \times 10^{-5}$ & " \\
\tableline
SiO& $ \rightarrow$ O$^+$ + Si & 678 & 1.3158$ \times 10^{-7}$ &1.0719$ \times 10^{-6}$ & 1.8973$ \times 10^{-6}$ & " \\
& $ \rightarrow$ Si$^+$ + O &218 & 4.0913$ \times 10^{-7} $&3.3327$ \times 10^{-6} $&5.8993$ \times 10^{-6} $& "  \\
& $ \rightarrow$ Si + O &110 & 8.0844$ \times 10^{-7} $&6.5855$ \times 10^{-6} $&1.1657$ \times 10^{-5} $& "  \\
& $ \rightarrow$ SiO$^+$ + e$^-$ &30 & 2.9722$ \times 10^{-6} $&2.4211$ \times 10^{-5} $&4.2857$ \times 10^{-5} $& "  \\
\tableline
N$_2$& $ \rightarrow$ N$^+$ + N & 264 & 3.3812$ \times 10^{-7}$ &2.7543$ \times 10^{-6}$ & 4.8755$ \times 10^{-6}$ & Khare \& Kumar (1977)\\
& $ \rightarrow$ N+ N &133.5 & 6.6813$ \times 10^{-7} $&5.4425$ \times 10^{-6} $&9.6339$ \times 10^{-6} $& "  \\
& $ \rightarrow$ N$_2^+$ + e$^-$ &36.3 & 2.4564$ \times 10^{-6} $&2.0009$ \times 10^{-5} $&3.5419$ \times 10^{-5} $& "  \\
\tableline
H & $ \rightarrow$ H$^+$ + e$^-$ &36.1& 2.4700$ \times 10^{-6} $&2.012$ \times 10^{-5} $&3.5615$ \times 10^{-5} $& Dalgarno Yan Liu (1999) \\
& $ \rightarrow$ H$^{\star}$ (n=2) &26.6 & 3.3521$ \times 10^{-6} $&2.7306$ \times 10^{-5} $&4.8335$ \times 10^{-5} $& "  \\
\tableline
He& $ \rightarrow$ He$^+$ + e$^-$ &46.3& 1.9258$ \times 10^{-6} $&1.5688$ \times 10^{-5} $&2.7769$ \times 10^{-5} $& "  \\
\tableline
H$_2$& $ \rightarrow$ H$^+$ + H &820 & 1.0874$ \times 10^{-7} $&8.8578$ \times 10^{-7} $&1.5679$ \times 10^{-6} $& "  \\
& $ \rightarrow$ H + H & 77 & 1.1580$ \times 10^{-6} $&9.433$ \times 10^{-6} $&1.6697$ \times 10^{-5} $& "  \\
& $ \rightarrow$ H$_2^+$ + e$^+$ &37.7 & 2.3651$ \times 10^{-6} $&1.9266$ \times 10^{-5} $&3.4103$ \times 10^{-5} $& "  \\
 \enddata
 \tablenotetext{a}{Arrhenius forms for k$_{C}$ (see text): {\rm A $\times \exp(-2976.5/T)$ (20 \Ms~progenitor) - A $\times \exp(-3464.1/T)$ (170 \Ms~progenitor) - A $\times \exp(-5376/T)$ (270 \Ms~progenitor)}}
\end{deluxetable}
\clearpage

\begin{deluxetable}{ccccccccccccc}
\tabletypesize{\scriptsize}
\rotate
\tablewidth{0pt}
\tablecaption{Initial (post-explosive) chemical composition in units of M$\odot$ for fully-mixed primordial SN ejecta without hydrogen mixing. The initial mean molecular weight $\mu_0{\rm(gas)}$ is given in g mole$^{-1}$.}
 \tablehead{
\colhead{M$_{prog.}$} & \colhead{$\mu_0$(gas)} & \colhead{He} & \colhead{O}&  \colhead{Si}& \colhead{S}& \colhead{Mg}& \colhead{Fe}& \colhead{C} & \colhead{Al}& \colhead{Ne}& \colhead{Ar}& \colhead{N}  
}
\startdata
20 M$\odot$& 5.63 & 3.59 &1.55&9.82 (-2)&4.12 (-2) &7.07 (-2) &7 (-2) &0.26 & 4.79 (-4)& 0.12& 6.9 (-3)& 2.69 (-4) \\
170 M$\odot$& 18.11 &1.96 & 44.23&16.16&8.66&1.94&3.63&2.30& 2.0 (-2)&1.19& 1.42& 1.0 (-2) \\
270 M$\odot$&  19.31 & 5.5&44.31&26.95&15.78&4.78&16.14&1.89&8.62 (-2)&4.70&2.60&1.26 (-2)  \\
\enddata
\end{deluxetable}
\clearpage
%
\begin{deluxetable}{lcccccccccccccc}
\rotate
\tabletypesize{\scriptsize}
\tablecaption{Initial (post-explosive) chemical composition in units of M$\odot$ for unmixed primordial SN ejecta. The initial mean molecular weight $\mu_0(gas)$ is given (in g mole$^{-1}$) as well as the C/O ratio of each zone.}
\tablewidth{0pt}
 \tablehead{
\colhead{Zone}& \colhead{$\mu_0$(gas)} & \colhead{C/O}& \colhead{He} & \colhead{O}&  \colhead{Si}& \colhead{S}& \colhead{Mg}& \colhead{Fe}& \colhead{C}& \colhead{Al}& \colhead{Cr}& \colhead{Co}& \colhead{Ni}  
}
\startdata
 {\bf Unmixed 20 M$_{\odot}$}& & & & & & & & & & & & & & \\
\tableline
Zone 1 (2.4-3 M$\odot$)&30.21&0.013& 0&6 (-4)&0.39&0.138&6 (-6)&0.048 &6 (-6) & 1.2 (-7) & 3 (-3) &4 (-4) & 2 (-4) \\
Zone 2 (3-3.6 M$\odot$)& 16.87 &0.0013& 0 &0.52&0.0358&0.0072&0.0363&0&4.98 (-4)&4.2 (-5)&0& 0&0  \\
Zone 3 (3.6-4.95 M$\odot$)&15.03 & 0.33 &0&1.08&2.7 (-5)&6.75 (-7)&4.725 (-3)&0&0.266&4.05 (-6)&0&0&0  \\
Zone 4 (4.95-5.85 M$\odot$)& 4.05 &29.47& 0.884&6.84 (-4)&9.0 (-9)&0&2.25 (-5)&0&1.512 (-2)&0&0&0&0  \\
\tableline
{\bf Unmixed 170 M$_{\odot}$}& & & & & & & & & & & & & & \\
\tableline
Zone 1 (0-20 M$\odot$)& 29.13& 0.066 &0&3.5 (-5)&13.2&4.0&5.3 (-5)&0.35&1.8 (-6) &1.8 (-10)& 3.5 (-3)& 1.2 (-3) & 3.5 (-3) \\
Zone 2 (20-40 M$\odot$)& 17.29 &2.9 $\times 10^{-5}$&0 &16.5&2.76&0.4&0.32&0&3.6 (-4)&5 (-4)&0& 0&0 \\
Zone 3 (40-55 M$\odot$)&  16.76 &0.03& 0&13.1&0.615&3 (-2)&1.22&0&3 (-2)&1.5 (-2)&0&0&0  \\
Zone 4 (55-78 M$\odot$)& 15.17 &0.29&  0&18.6&1.84 (-3)&1.38 (-6)&0.299&0&4.07&4.6 (-4)&0&0&0 \\
Zone 5 (78-82 M$\odot$)& 10.46& 0.56 & 0.596&2.4&4 (-6)&2.4 (-7)&3.6 (-3)&0&1.0&4 (-8)&0&0&0  \\
\enddata
\end{deluxetable}
\clearpage

\begin{deluxetable}{lcccc}
\tabletypesize{\scriptsize}
\tablecaption{Mass yields of most important molecules ejected at day 1000 for fully-mixed ejecta with and without hydrogen mixing. \tablenotemark{a,b} }
 \tablewidth{0pt}
 \tablehead{
\colhead{ Molecules } & \colhead{170 M$_{\odot}$ } & \colhead{170 M$_{\odot}$ } & \colhead{270 M$_{\odot}$ } & \colhead{20 M$_{\odot}$\tablenotemark{c}  } \\
\colhead{   } & \colhead{no hydrogen} & \colhead{10 \% hydrogen} & \colhead{no hydrogen} & \colhead{no hydrogen} \\
}
\startdata
CO & 5.61& 0.18 & 3.22 & 0.63 \\
SO & 5.61 &13.28 & 1.68 $\times 10^{-4}$&  6.35 $\times 10^{-2}$   \\
O$_2$&  5.24 $\times 10^{-2}$ & 24.71& 2.67 $\times 10^{-4}$ & 8.79 $\times 10^{-2}$ \\ 
CO$_2$ & 7.20 $\times 10^{-5}$ & 8.49 &3.45 $\times 10^{-8}$ &1.48$\times 10^{-4}$  \\
H$_2$ &0& 0.12& 0 & 0\\
N$_2$ & 1.03 $\times 10^{-2}$ & 3.42 $\times 10^{-3}$ &5.70 $\times 10^{-4}$ & 2.73 $\times 10^{-4}$ \\
NO & 5.21 $\times 10^{-4}$  &1.62 $\times 10^{-3}$ & 1.88 $\times 10^{-7}$ & 2.47 $\times 10^{-6}$  \\
OH &0&1.31 $\times 10^{-3}$ &0&0 \\
\tableline
Total& {\bf 11.29}   & {\bf 46.80} & {\bf 3.22 }& {\bf 0.78 } \\
Efficiency&  {\bf 13.71 \% }& {\bf 56.86 \%}&{\bf 2.49\%}&{\bf 13.42  \%} \\

 \enddata
\tablenotetext{a}{Mass yields are in \Ms}
\tablenotetext{b}{The efficiency is defined as the ratio of the molecular mass to the He core mass} 
\tablenotetext{c}{A mass cut of 2.4 \Ms~is assumed as in Nozawa et al. (2003)}
\end{deluxetable}

\clearpage

\begin{deluxetable}{lcccccl}
\tabletypesize{\scriptsize}
\tablecaption{Mass yields of most important molecules ejected at day 1000 for the unmixed ejecta of the 170 \Ms~progenitor without hydrogen mixing. \tablenotemark{a,b}}
 \tablewidth{0pt}
 \tablehead{
\colhead{} & \colhead{Zone 1 } & \colhead{Zone 2 } & \colhead{Zone 3 } & \colhead{Zone 4 } & \colhead{Zone 5 } & \colhead{Zone 1-5} \\
\colhead{ Zone mass}& \colhead{(20 M$_{\odot}$)} & \colhead{(20 M$_{\odot}$)} & \colhead{(15 M$_{\odot}$)} & \colhead{(23 M$_{\odot}$)} & \colhead{(4.3 M$_{\odot}$)} & \colhead{(82.3 M$_{\odot}$)}\\
\colhead{Zone C/O } & \colhead{0.066 } & \colhead{2.9$\times 10^{-5}$ } & \colhead{0.03 } & \colhead{0.29 } & \colhead{0.56 } & \colhead{} 
}
\startdata
O$_2$ & 0& 6.24& 5.98 & 13.80 & 9.28 $\times 10^{-5}$&  26.00\\
SiS & 7.36 &0 & 0&  8.79 $\times 10^{-6}$  & 0  &  7.36 \\
CO& 0 & 8.39$\times 10^{-4}$ & 6.98 $\times 10^{-2}$ & 9.95 $\times 10^{-1}$ &2.03 &  3.09\\ 
SO & 9.06$\times 10^{-5}$   & 5.97$\times 10^{-1}$  &4.47 $\times 10^{-2}$ &0 & 0 & 0.64\\
\tableline
Total mass & 7.36  &6.84 & 6.10 &14.80  & 2.03  &{\bf 37.09}  \\
Efficiency & 41.81\% &34.18 \%&  40.64 \%& 64.22 \% &47.21 \% & {\bf 45.07 \%}  \\
 \enddata
 \tablenotetext{a}{Mass yields are in \Ms}
 \tablenotetext{b}{The efficiency is defined as the ratio of the molecular mass to the zone mass}
\end{deluxetable}
\clearpage

\begin{deluxetable}{lccccl}
\tabletypesize{\scriptsize}
\tablecaption{Mass yields of most important species ejected at day 1000 for the unmixed ejecta of the 20 \Ms~progenitor without hydrogen mixing. \tablenotemark{a,b}}
 \tablewidth{0pt}
 \tablehead{
\colhead{  } & \colhead{Zone 1} & \colhead{Zone 2} & \colhead{Zone 3} & \colhead{Zone 4} & \colhead{Zones 1- 4} \\
\colhead{ Zone mass}& \colhead{(0.6 M$_{\odot}$)\tablenotemark{c}} & \colhead{( 0.6 M$_{\odot}$)} & \colhead{(1.35 M$_{\odot}$)} & \colhead{(0.9 M$_{\odot}$)} &  \colhead{(3.45 M$_{\odot}$)}\\
\colhead{Zone C/O } & \colhead{0.013 } & \colhead{0.0013 } & \colhead{0.33 } & \colhead{29.47 } & \colhead{}
}
\startdata
O$_2$ & 0& 0.29& 0.45 & 0 & 0.74\\
CO& 0 & 1.15$\times 10^{-3}$ & 0.27 & 3.63 $\times 10^{-5}$ &0.27 \\ 
SiS & 0.21 &0 & 0&  0  & 0.21   \\
SO & 9.43$\times 10^{-4}$   & 1.03$\times 10^{-2}$  &9.49 $\times 10^{-7}$ &0 & 0.01 \\
\tableline
Total mass & 0.21  & 0.3 & 0.72 &0   & {\bf 1.23 } \\
Efficiency & 35.37\% &50.14\%&  53.41 \%&0.004 \% & {\bf 35.75 \%} \\ 
\enddata
\tablenotetext{a}{Mass yields are in \Ms}
 \tablenotetext{b}{The efficiency is defined as the ratio of the molecular mass to the zone mass}
 \tablenotetext{c}{A mass cut of 2.4 \Ms~is assumed for Zone~1 as in Nozawa et al. (2003)}
\end{deluxetable}
\clearpage


\begin{figure}
\epsscale{1.15}
\plottwo{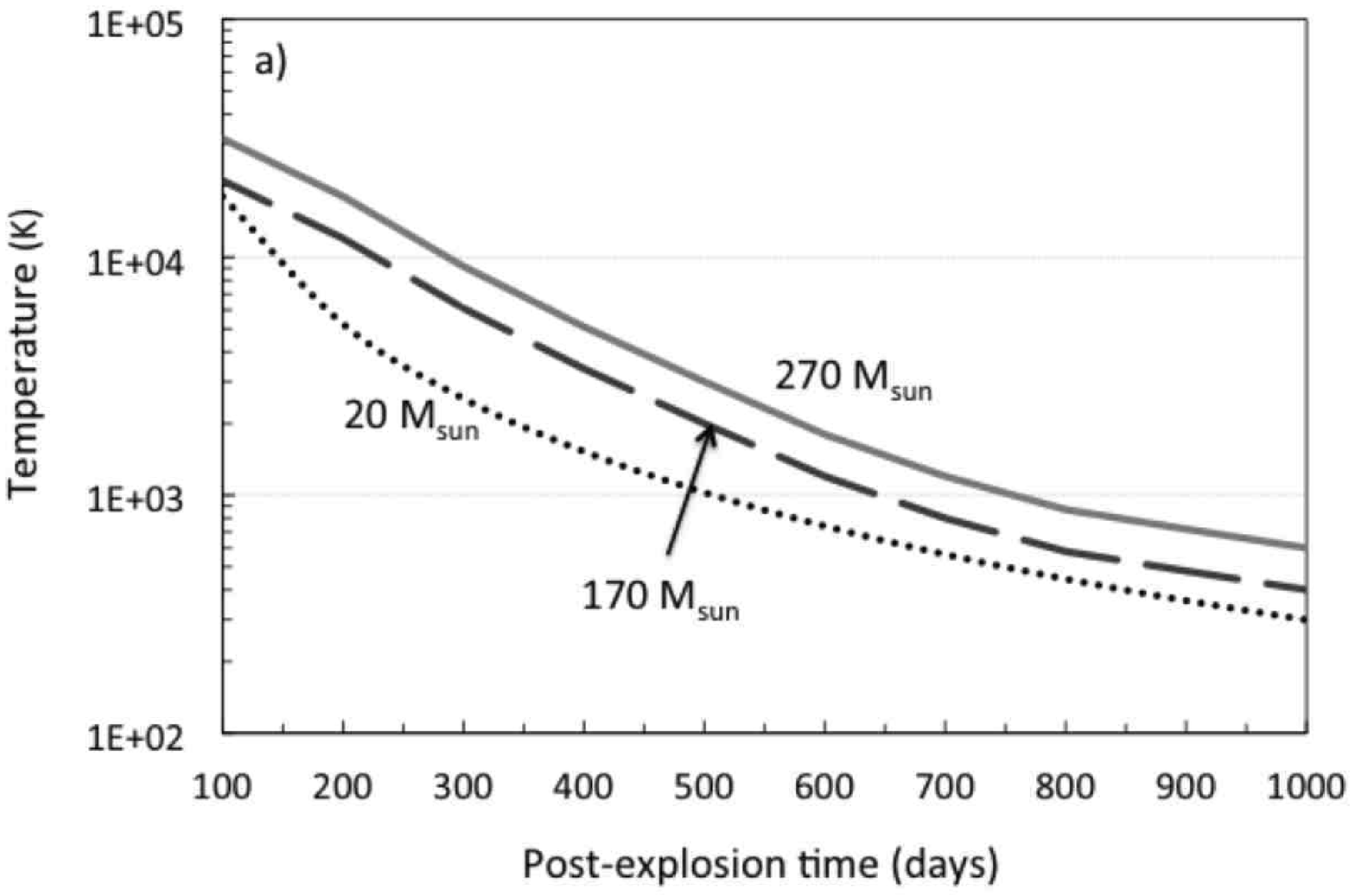}{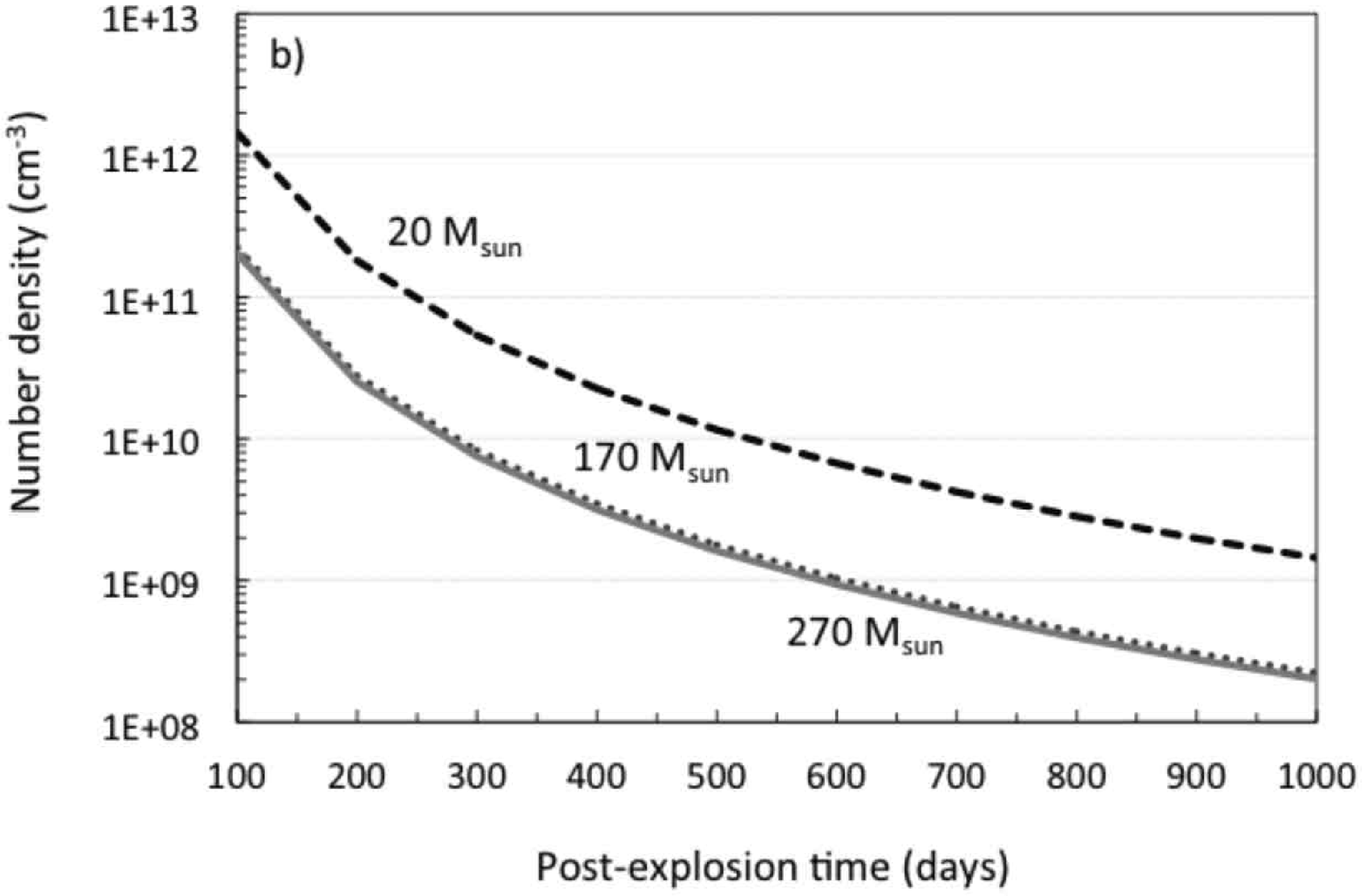}
\caption{The evolution of the gas parameters as a function of time for the different SN ejecta (taken from Nozawa et al. 2003) and assumed to be independent of mass zone within the He core. a) Temperature where the 270 \Ms~profile has been rescaled by a factor 1.5 compared to that of the 170 \Ms~ejecta to account for the larger explosion energy; b) Number density.}
\end{figure}


\clearpage
\begin{figure}
\epsscale{0.8}
\plotone{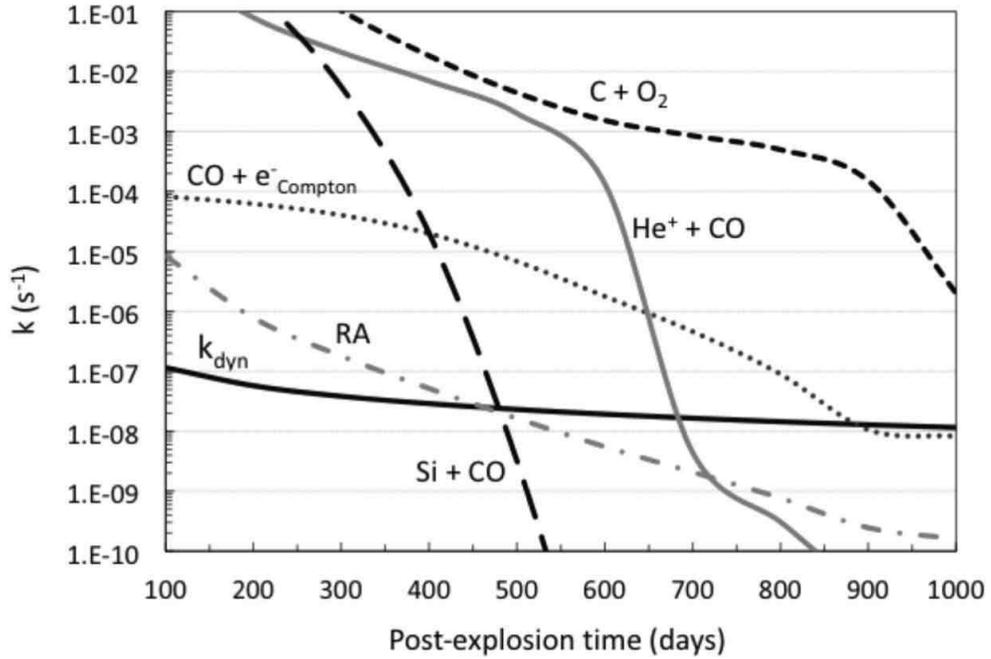}
\caption{The time dependence of the rates of the major processes leading to the formation of CO (eqs. (\ref{co})-(\ref{coe}) is compared to the inverse of the dynamical time scale of the ejecta $k_{dyn}$ for the fully mixed 170 \Ms~case without hydrogen diffusion. The figure shows when reactions proceeds and freeze out, leading to an active chemistry far from Steady-State.}
\end{figure}


\clearpage
\begin{figure}
\epsscale{1.15}
\plottwo{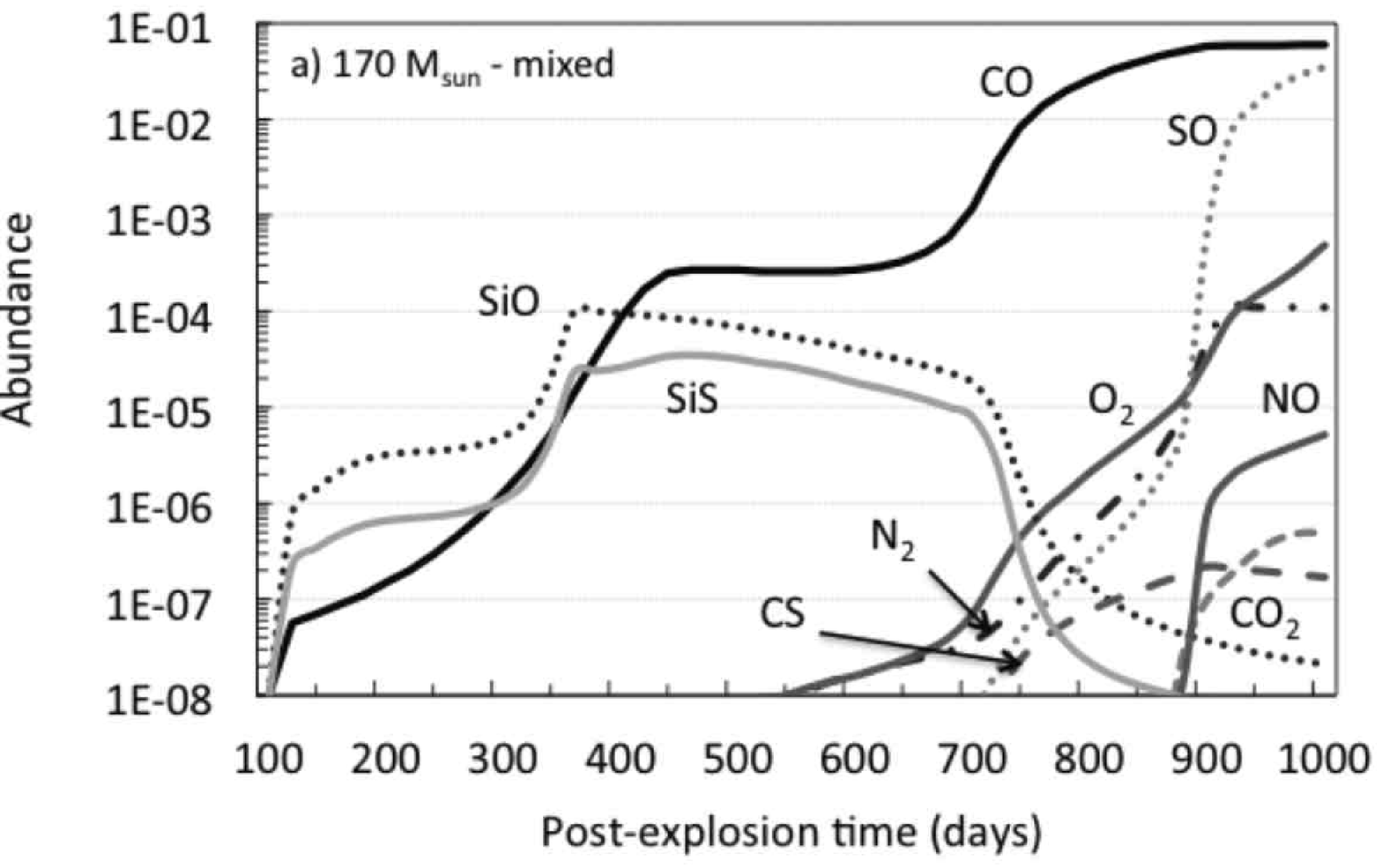}{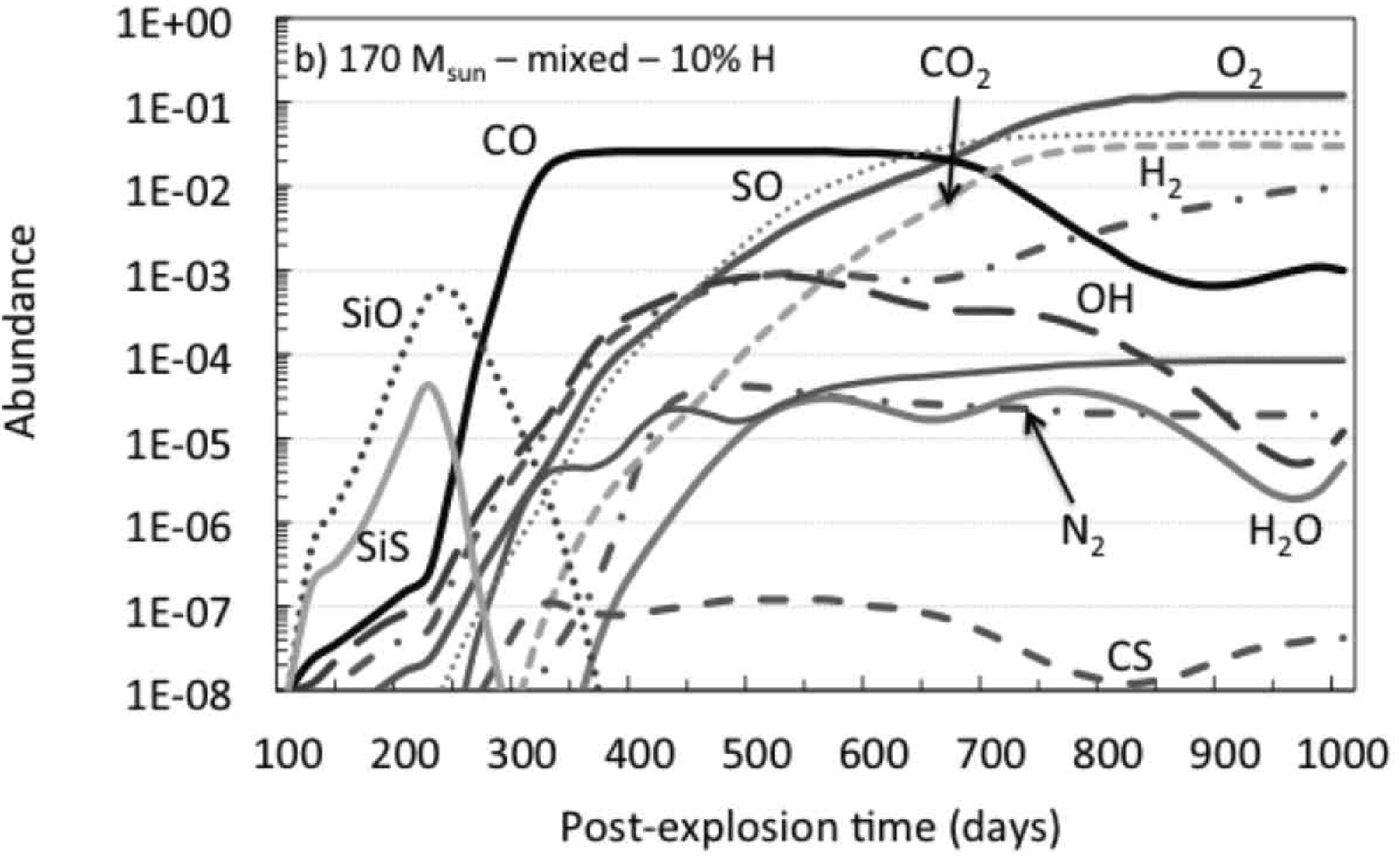}
\caption{The evolution of molecular abundances normalized to total gas number density for the fully mixed ejecta of the 170 M$_{\odot}$ progenitor when no hydrogen mixing is considered (a), and when 10 \% of the hydrogen mass of the progenitor envelope is microscopically mixed to the helium core (b). The figure illustrates the sensitivity of the chemistry to the presence of hydrogen in the ejecta.}
\end{figure}


\clearpage
\begin{figure}
\epsscale{1.15}
\plottwo{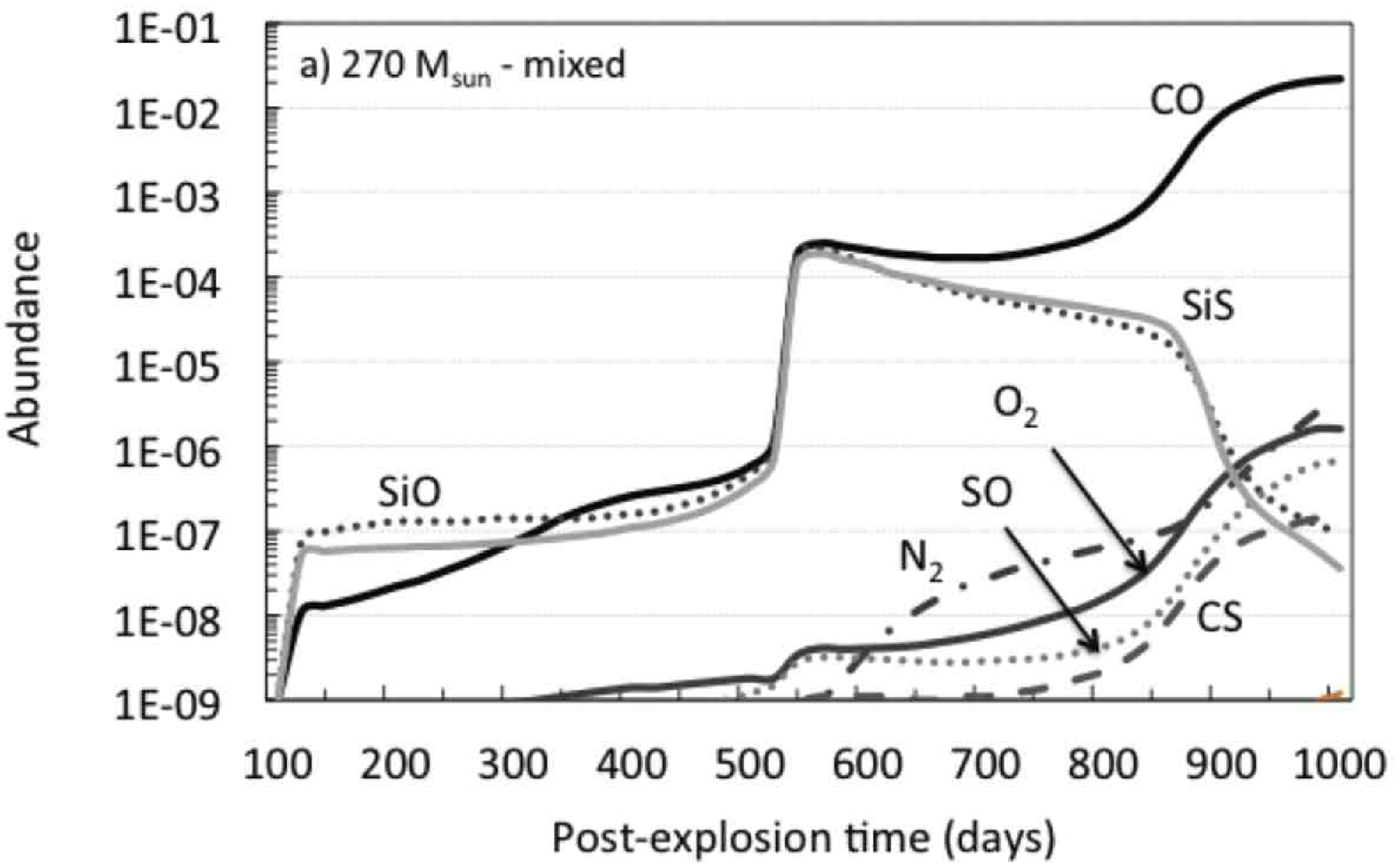}{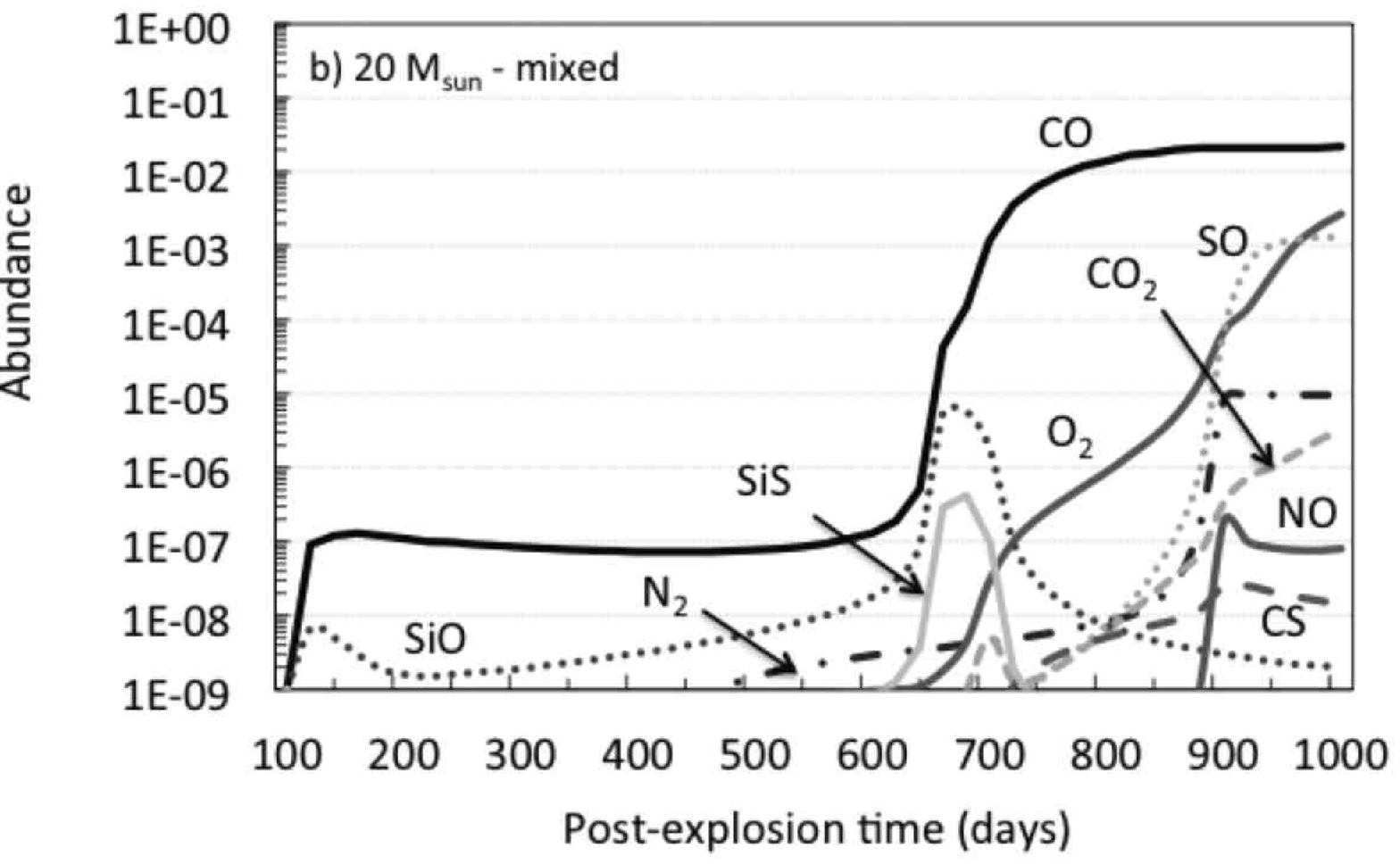}
\caption{The evolution of the molecular abundances normalized to total gas number density for the fully mixed ejecta: a) 270 M$_{\odot}$ progenitor; b) 20 M$_{\odot}$  progenitor. No hydrogen mixing is considered in both cases.}
\end{figure}


\clearpage
\begin{figure}
\epsscale{1.15}
\plottwo{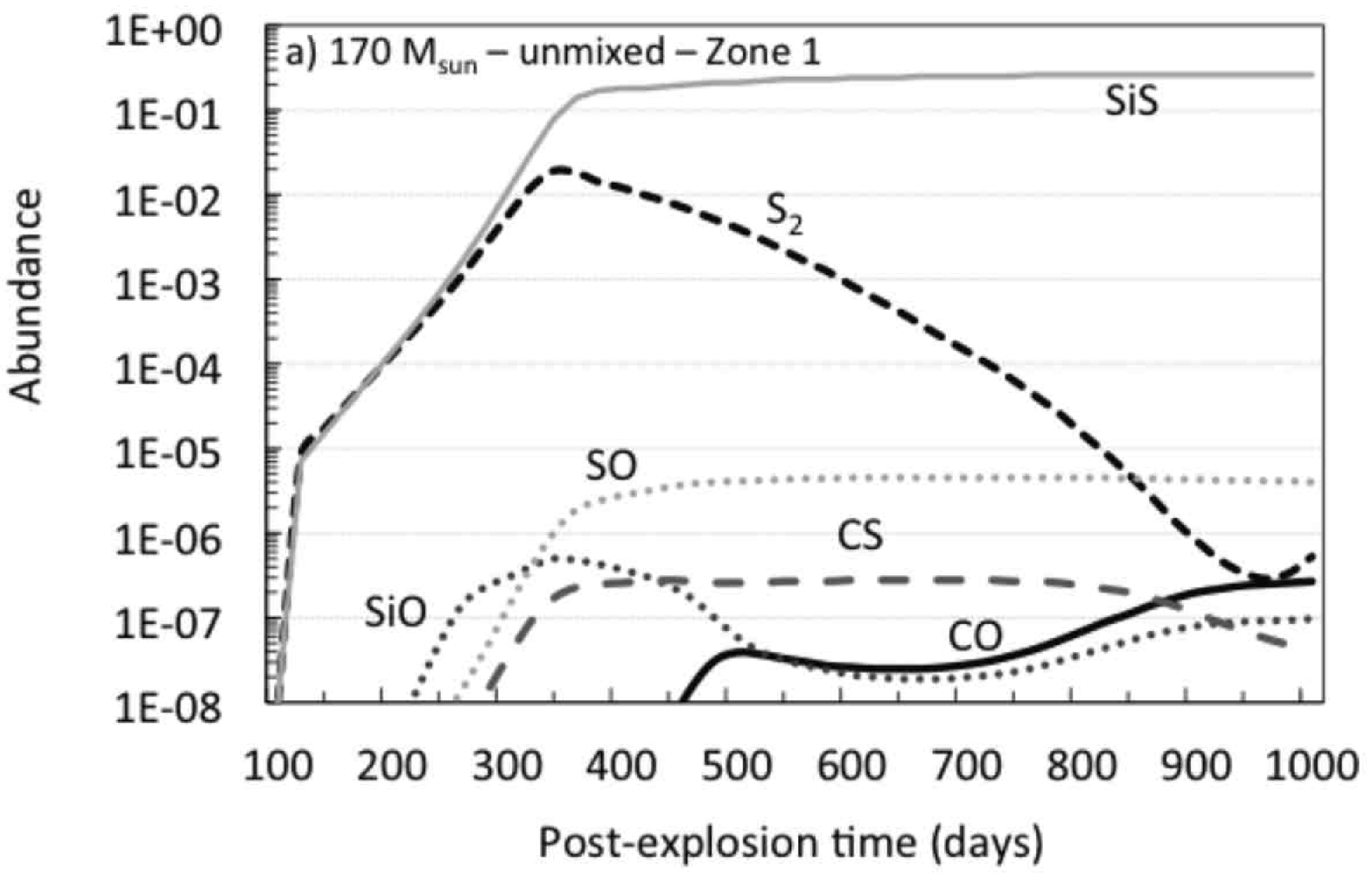}{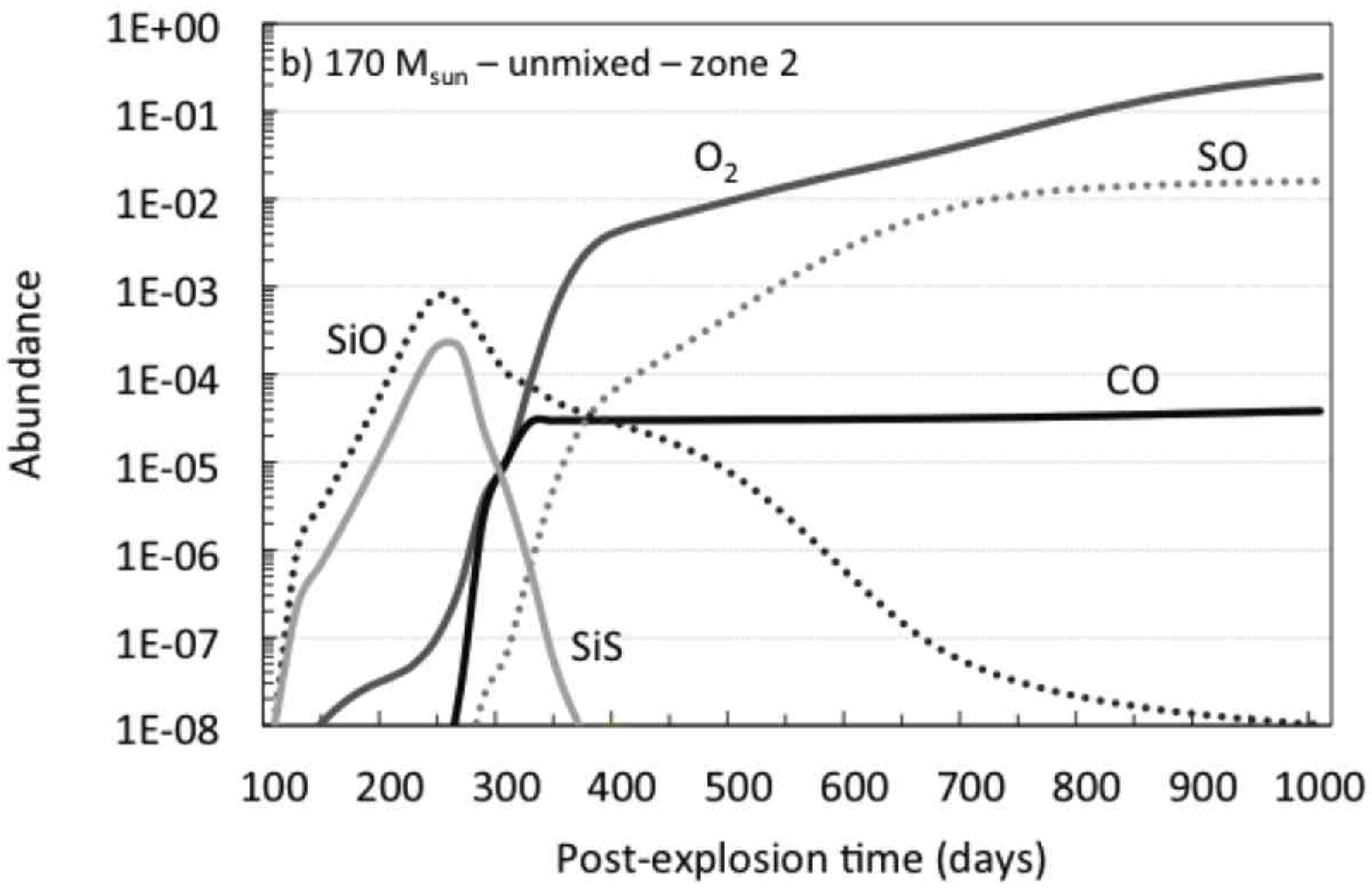}
\caption{The evolution of the molecular abundances normalized to total gas number density for the 170 M$_{\odot}$ unmixed ejecta: a) The Si/S/Fe-rich zone 1;  b) The O/Si/Mg/S-rich zone 2.}
\end{figure}


\clearpage
\begin{figure}
\epsscale{1.15}
\plottwo{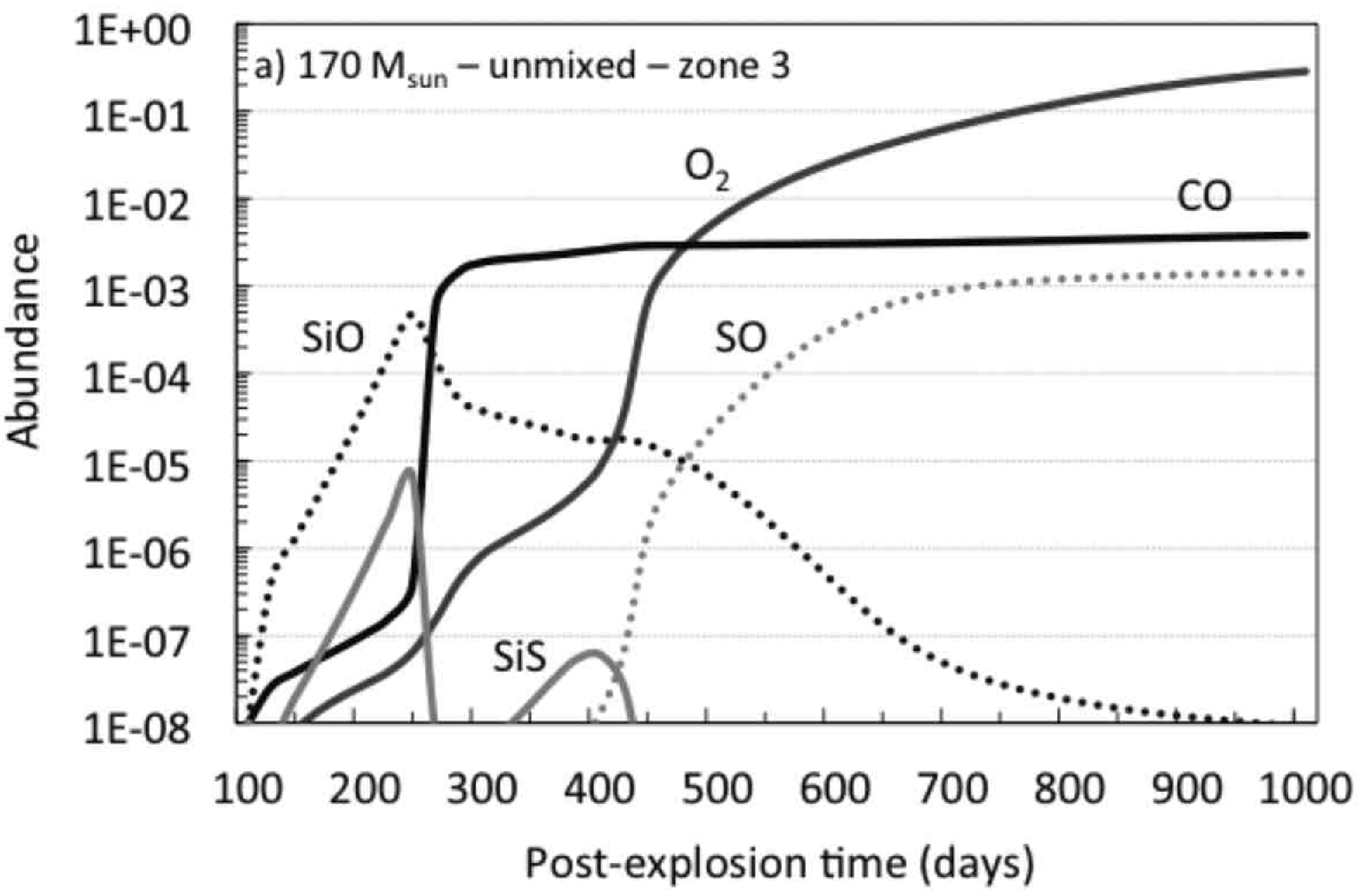}{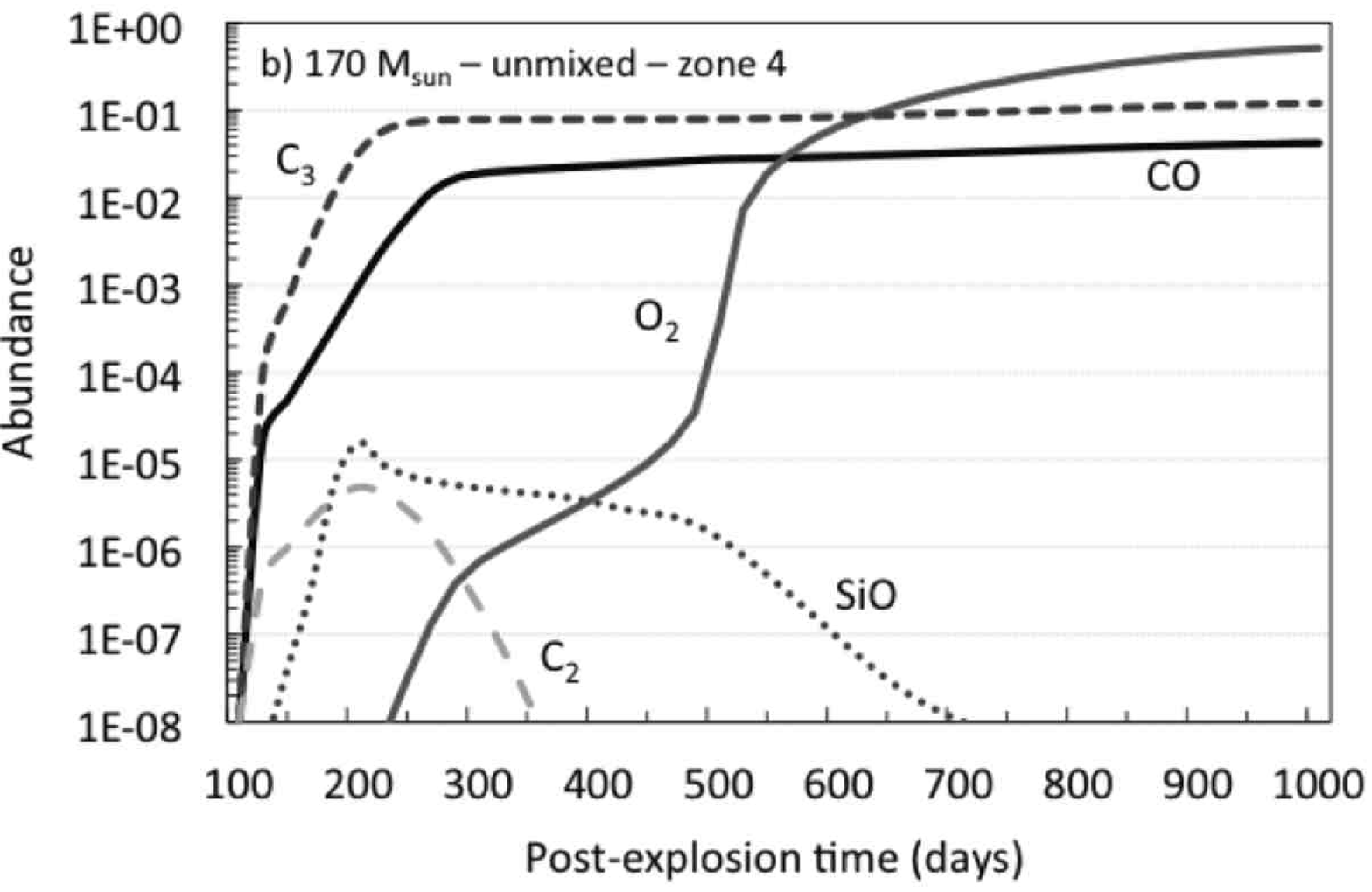}
\caption{The evolution of the molecular abundances normalized to total gas number density for the 170 M$_{\odot}$ unmixed ejecta: a) The O/Mg/Si-rich zone 3;  b) The O/C/Mg-rich zone 4 in which the rapid conversation of CO to C$_2$ take place despite a C/O ratio of 0.29. }
\end{figure}


\clearpage
\begin{figure}
\epsscale{0.8}
\plotone{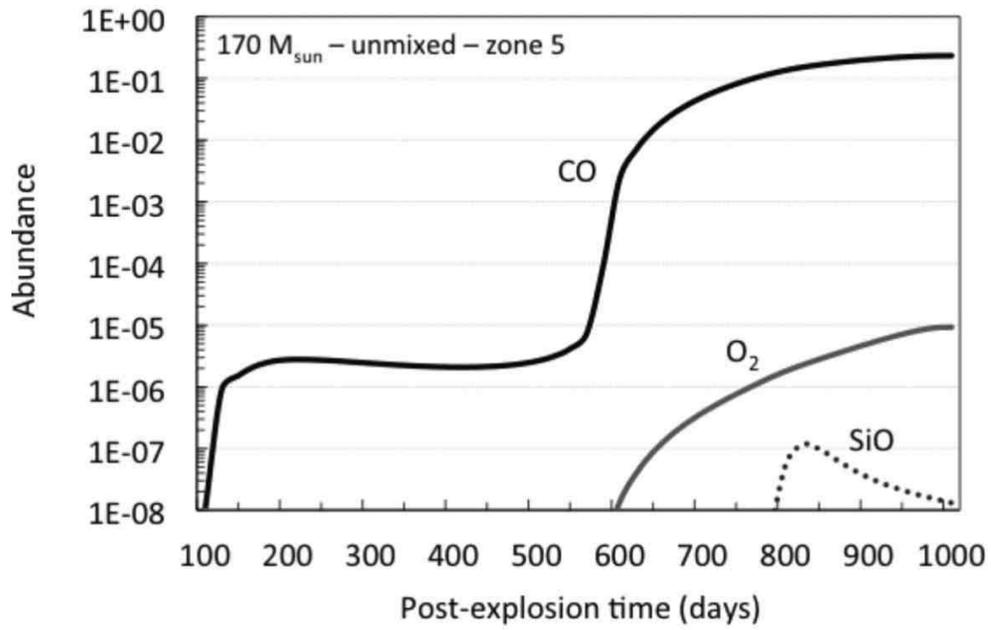}
\caption{The evolution of the molecular abundances normalized to total gas number density in the O/C/He-rich zone 5 of the 170 M$_{\odot}$ unmixed ejecta. CO is the only molecule to form due to the destruction of molecules by He$^+$.}
\end{figure}


\clearpage
\begin{figure}
\epsscale{0.8}
\plotone{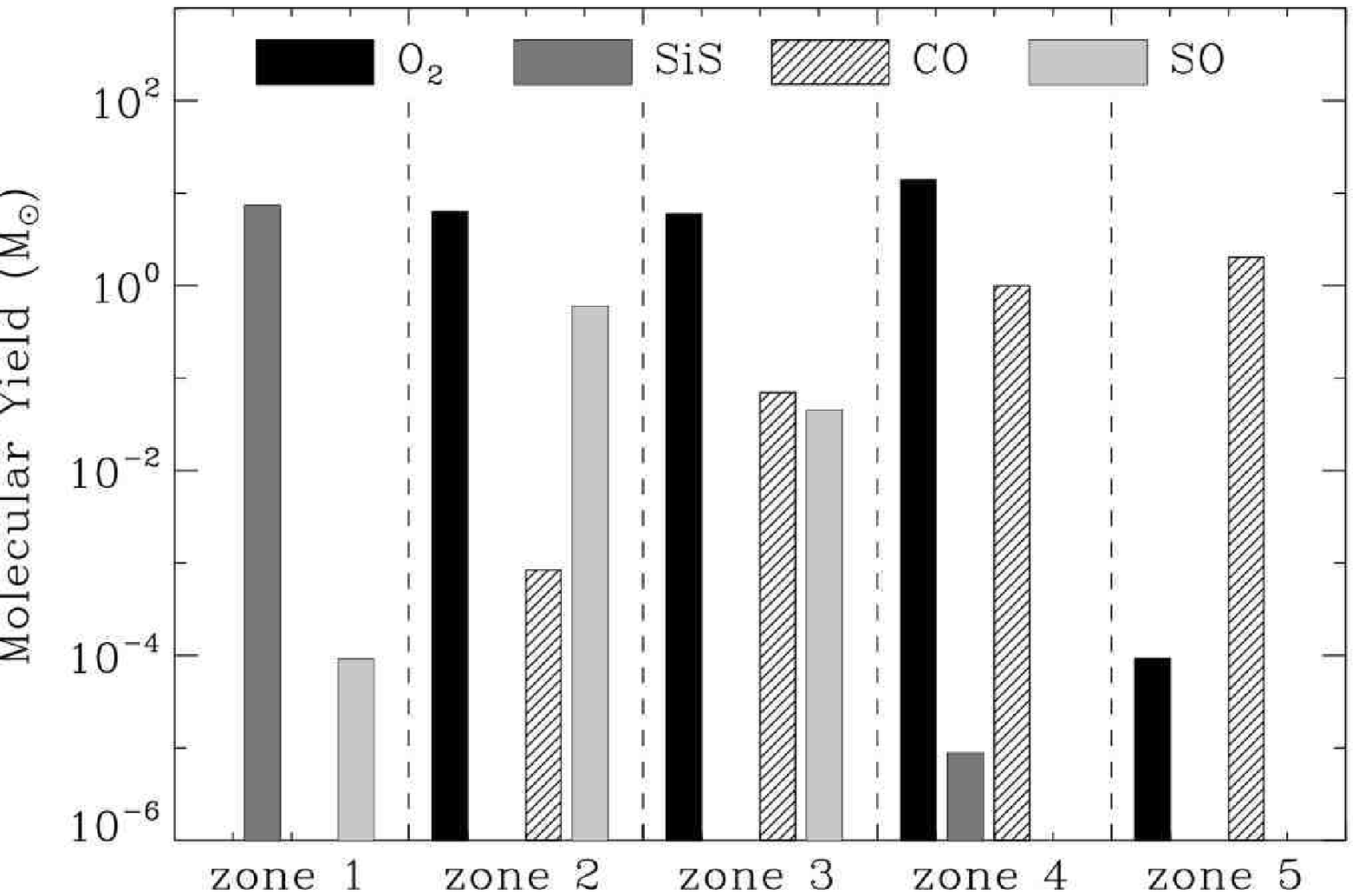}
\caption{The mass yield (in \Ms) of molecules ejected at day 1000 for the unmixed ejecta of the 170 \Ms~progenitor.   }
\end{figure}


\clearpage
\begin{figure}
\epsscale{0.8}
\plotone{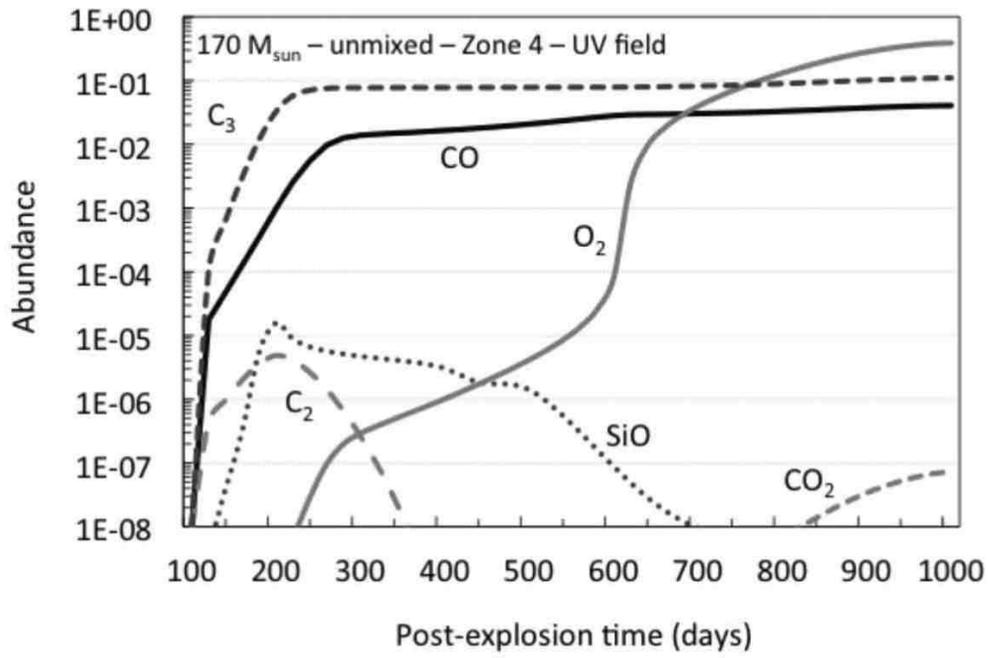}
\caption{The evolution of the molecular abundances normalized to total gas number density for zone 4 of the 170 M$_{\odot}$ unmixed ejecta where UV radiation is included.  Molecular abundances are quasi similar to those in Figure 6 except for O$_2$ whose formation is delayed to later times due to UV destruction. }
\end{figure}


\clearpage
\begin{figure}
\epsscale{1.15}
\plottwo{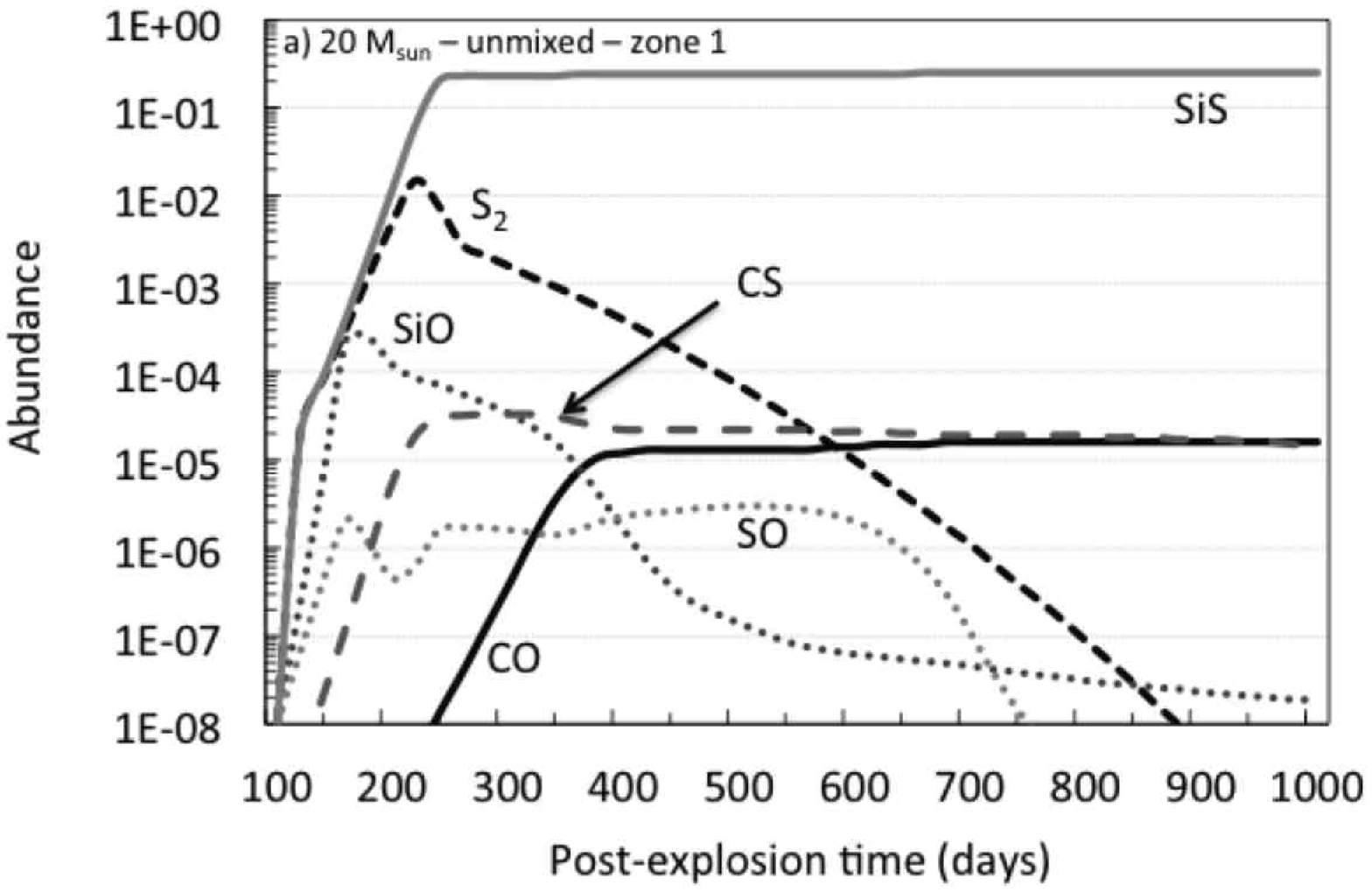}{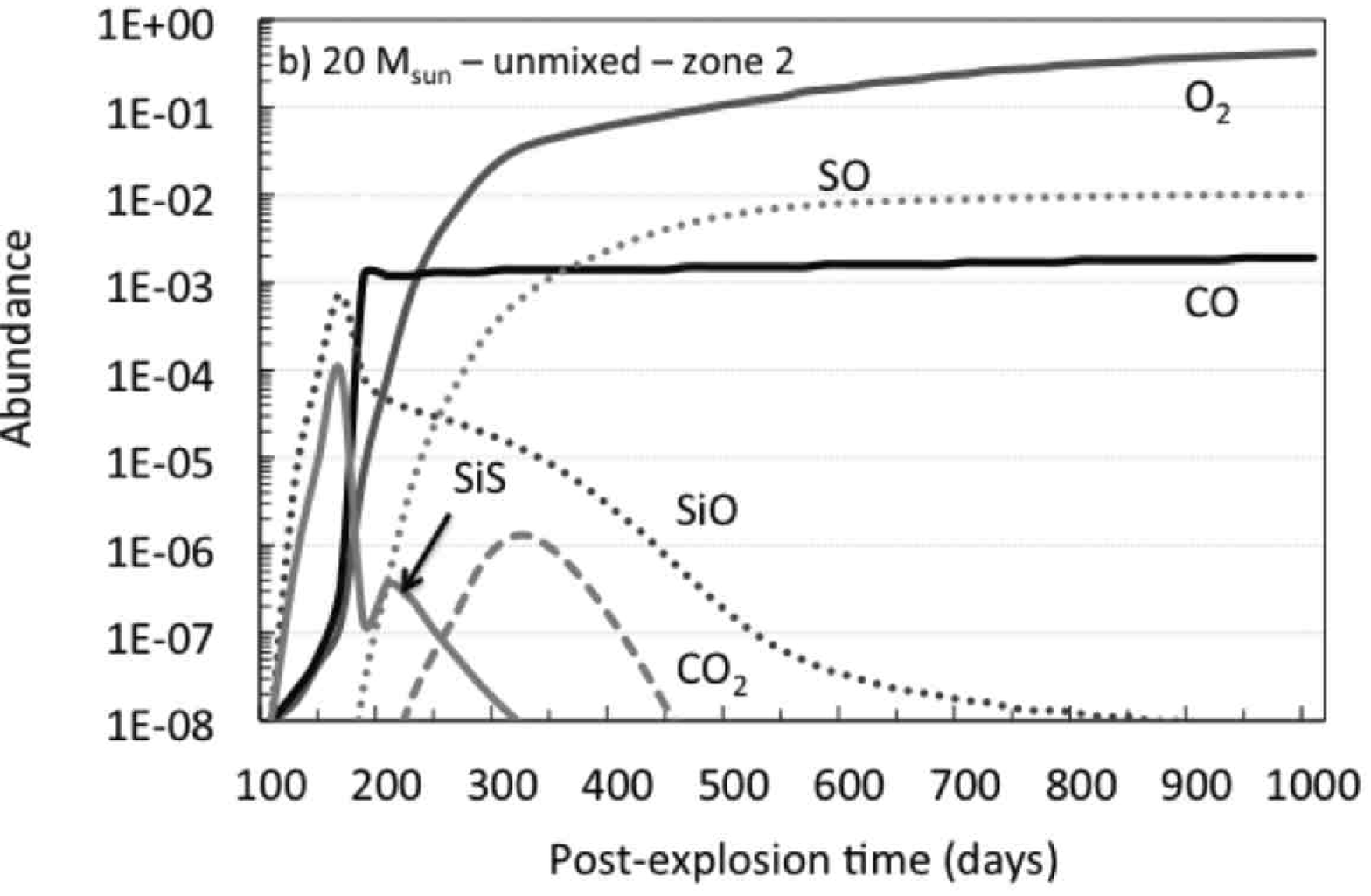}
\caption{The evolution of the molecular abundances normalized to total gas number density for the 20 M$_{\odot}$ unmixed ejecta: a) The Si/S/Fe-rich zone 1;  b) The O/Si/Mg/S-rich zone 2.}
\end{figure}


\clearpage
\begin{figure}
\epsscale{0.8}
\plotone{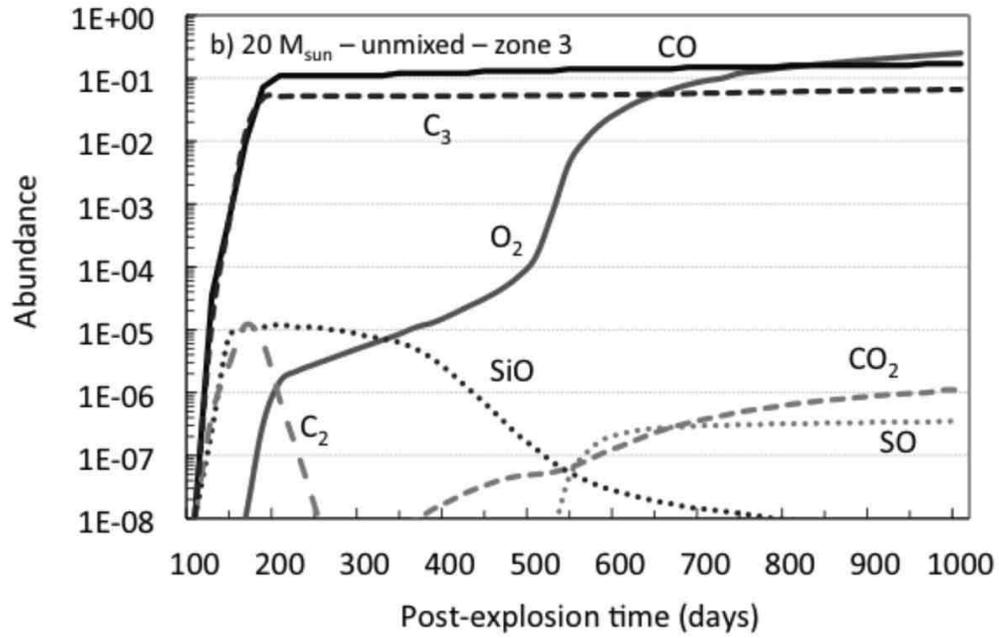}
\caption{The evolution of the molecular abundances normalized to total gas number density for the O/C/Mg-rich zone 3 of the 20 M$_{\odot}$ unmixed ejecta. The rapid conversation of CO to C$_2$ takes place despite a C/O ratio of 0.33. }
\end{figure}
\clearpage


\clearpage
\begin{figure}
\epsscale{0.8}
\plotone{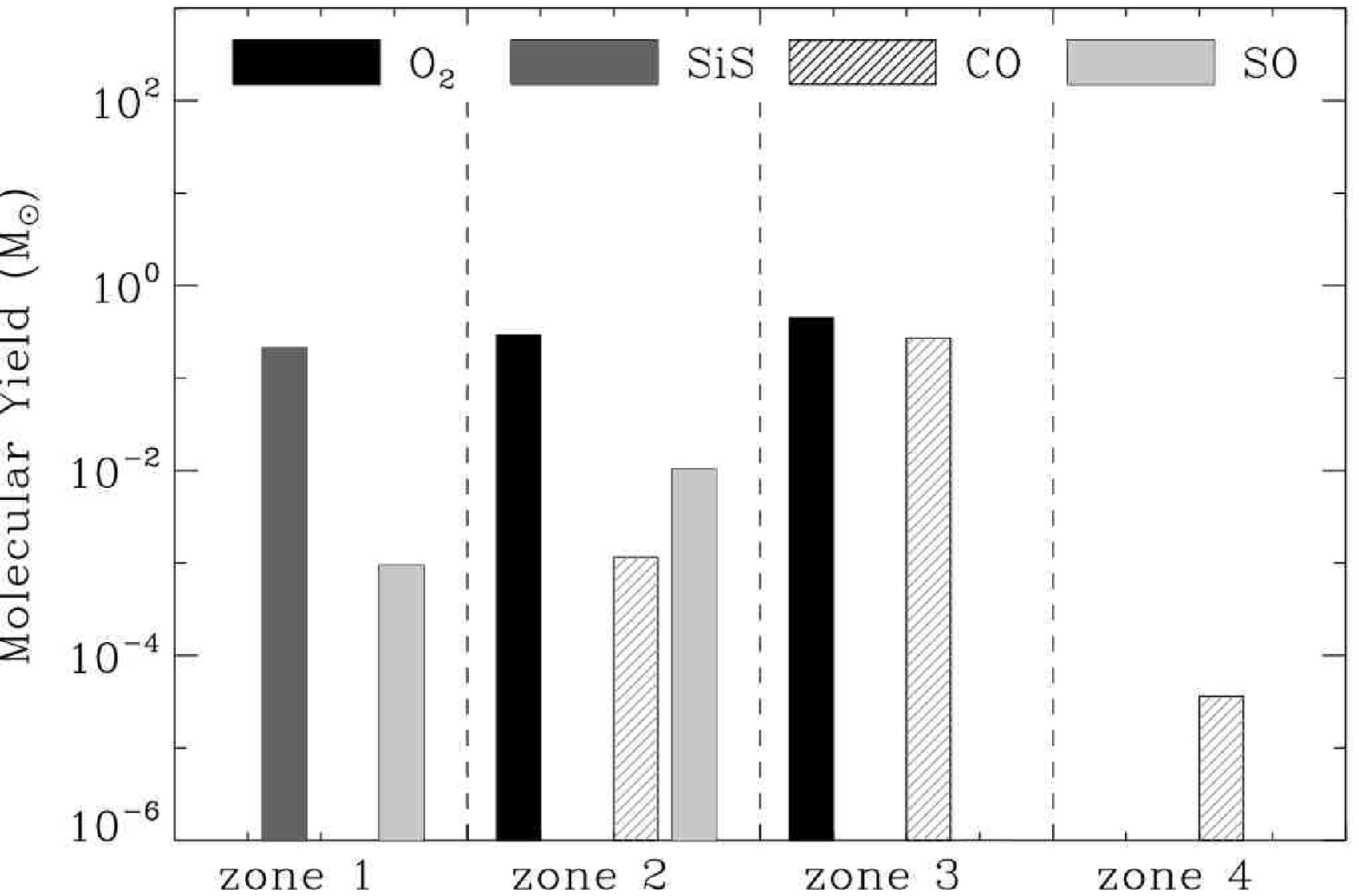}
\caption{The mass yield (in \Ms) of molecules ejected at day 1000 for the unmixed ejecta of the 20 \Ms~progenitor. }
\end{figure}


\appendix

\begin{deluxetable}{llclrrrr}
\tabletypesize{\scriptsize}
\tablecaption{Chemical processes considered in the present study. Chemical reaction rate coefficients are given according to equation (\ref{kij}). The Compton electron destruction and UV photodissociation reactions are listed for the 170~\Ms~progenitor. Rate values for other progenitors can be derived from Table 3.  }
 \tablewidth{0pt}
 \tablehead{
\colhead{ } &\multicolumn{3}{c}{Chemical processes}  & \colhead{A } & \colhead{$\nu$} & \colhead{E$_a$} & \colhead{Reference \tablenotemark{a}}
}
\startdata
\multicolumn{8}{c}{\bf TERMOLECULAR}  \\
\tableline
3B1 &H + H + M &$\longrightarrow$& H$_2$ + M & 6.84$\times 10^{-33}$& -1& 0 & NIST\\
3B2 & H + C + M& $\longrightarrow$ &CH + M & 1.00$\times 10^{-33}$& 0& 0 & E\\
3B3 &H + O + M &$\longrightarrow$ &OH + M & 4.36$\times 10^{-32}$& -1& 0 & NIST\\
3B4 &H + OH + M& $\longrightarrow$& H$_2$O + M & 2.59$\times 10^{-31}$& -2& 0 & NIST\\
3B5 &H + CN + M &$\longrightarrow$ &HCN + M & 8.63$\times 10^{-30}$& -2.2& 566.5 & NIST\\
3B6 &H + CO + M &$\longrightarrow$ &HCO+ M & 5.29$\times 10^{-34}$& 0& 370.4 & NIST \\
3B7 &H + C$_2$ + M& $\longrightarrow$& C$_2$H+ M & 1.00$\times 10^{-33}$& 0& 0& E \\
3B8 &H + C$_3$ + M &$\longrightarrow$ &C$_3$H+ M & 1.00$\times 10^{-33}$& 0& 0& E \\
3B9 &O + C + M &$\longrightarrow$ &CO + M & 2.14$\times 10^{-29}$& -3.08& -2114.0 & UDFA06\\
3B10 &O + O + M &$\longrightarrow$ &O$_2$ + M & 9.26$\times 10^{-34}$& -1& 0& NIST\\
3B11 &O + S + M &$\longrightarrow$ &SO+ M & 9.26$\times 10^{-34}$& -1& 0& E as 3B9\\
3B12&O +N + M &$\longrightarrow$ &NO+ M & 5.46$\times 10^{-33}$& 0& 155.1& NIST\\
3B13 &O + CO + M &$\longrightarrow$& CO$_2$+ M & 1.20$\times 10^{-32}$& 0& 2160.0 & NIST\\
3B14 &O + Si + M &$\longrightarrow$ &SiO+ M & 2.14$\times 10^{-29}$& -3.08& -2114.0 & E as 3B9\\
3B15 &C + C + M &$\longrightarrow$ &C$_2$+ M & 1.00$\times 10^{-33}$& 0& 0& E \\
3B16 &C + C$_2$ + M &$\longrightarrow$ &C$_3$+ M & 1.00$\times 10^{-33}$& 0& 0& E \\
3B17 &C + S + M &$\longrightarrow$ &CS+ M & 2.14$\times 10^{-29}$& -3.08& -2114.0 & E as 3B9\\
3B18 &C + N + M &$\longrightarrow$ &CN+ M & 9.40$\times 10^{-33}$& 0& 0 & NIST\\
3B19 &Si + H + M &$\longrightarrow$ &SiH+ M & 1.00$\times 10^{-33}$& 0& 0 & E \\
3B20 &Si + N + M &$\longrightarrow$ &SiN+ M & 9.40$\times 10^{-33}$& 0& 0 & E as 3B17 \\
3B21 &Si + S + M &$\longrightarrow$ &SiS+ M & 2.14$\times 10^{-29}$& -3.08& -2114.0 & E as 3B9 \\
3B22 &S + S + M &$\longrightarrow$ &S$_2$+ M & 2.76$\times 10^{-33}$& 0& 0& NIST \\
3B23 &N+ N + M &$\longrightarrow$ &N$_2$+ M & 1.25$\times 10^{-32}$& 0& 0& NIST\\
3B24&CO + CH + M &$\longrightarrow$& HC$_2$O+ M & 2.80$\times 10^{-34}$& -0.4& 0& NIST\\
3B25 &H + CO + H &$\longrightarrow$ &O+ CH$_2$ & 1.00$\times 10^{-33}$& 0& 0& E \\
3B26 &H + CO$_2$ + H &$\longrightarrow$ &O$_2$+ CH$_2$ & 1.00$\times 10^{-33}$& 0& 0& E \\
3B27 &H + C$_2$H$_2$ + H &$\longrightarrow$& CH$_2$+ CH$_2$ & 1.00$\times 10^{-36}$& 0& 0& E \\
3B28&CO + OH + H &$\longrightarrow$& O$_2$+ CH$_2$ & 1.00$\times 10^{-33}$& 0& 0& E\\
\tableline
\multicolumn{8}{c}{\bf THERMAL FRAGMENTATION}  \\
\tableline
TF1 &H$_2$ + M &$\longrightarrow$& H + H+M & 2.54$\times 10^{-8}$& -0.1& 52555.6 & NIST\\
TF2 &CH + M &$\longrightarrow$ &C + H+M & 3.16$\times 10^{-10}$& 0& 33700.0 & NIST\\
TF3 &OH+ M &$\longrightarrow$ &O + H+M & 4.00$\times 10^{-9}$& -0.1& 50000.0 & NIST\\
TF4 &SiH + M &$\longrightarrow$ &Si+ H+M & 3.16$\times 10^{-10}$& 0& 33700.0& E as TF2\\
TF5 &H$_2$O + M &$\longrightarrow$ &OH + H+M & 5.80$\times 10^{-9}$& 0& 52920.0 & NIST\\
TF6 &HCN+ M &$\longrightarrow$ &H+CN+M &2.08$\times 10^{-8}$& 0& 54630.0& NIST\\
TF7 &CH$_2$+ M &$\longrightarrow$ &CH+H+M &6.64$\times 10^{-9}$& 0& 41852.0& NIST\\
TF8 &CH$_2$+ M &$\longrightarrow$ &H$_2$+C+M &2.66$\times 10^{-10}$& 0& 32230.0& NIST\\
TF9 &C$_2$H+ M &$\longrightarrow$ &C$_2$+H+M &3.75$\times 10^{-10}$& 0&50040.0& NIST\\
TF10 &C$_3$H+ M &$\longrightarrow$ &C$_3$+H+M &1.00$\times 10^{-10}$& 0&48600.0& E\\
TF11 &HC$_2$O+ M &$\longrightarrow$ &CO+CH+M &1.08$\times 10^{-8}$& 0&29585.0& NIST\\
TF12 &O$_2$+ M &$\longrightarrow$ &O+O+M &5.17$\times 10^{-10}$& 0& 58410.0 & NIST\\
TF13 &CO+ M &$\longrightarrow$ &C+O+M &4.40$\times 10^{-10}$& 0& 98600.0 & Appleton 1970\\
TF14 &SO+ M &$\longrightarrow$ &S+O+M &6.61$\times 10^{-10}$& 0& 53885.0 & NIST\\
TF15 &SiO+ M &$\longrightarrow$ &Si+O+M &4.40$\times 10^{-10}$& 0& 98600.0 & E as TF13\\
TF16 &NO+ M &$\longrightarrow$ &N+O+M &4.10$\times 10^{-9}$& 0& 75380.0 & NIST\\
TF17 &CO$_2$+ M &$\longrightarrow$ &CO+O+M &8.02$\times 10^{-11}$& 0& 26900.0 & NIST\\
TF18 &C$_2$+ M &$\longrightarrow$ &C+C+M &3.02$\times 10^{-9}$& 0& 63980.7 & NIST\\
TF19&CS+ M &$\longrightarrow$ &C+S+M &4.40$\times 10^{-10}$& 0& 98600.0 & E as TF13\\
TF20&CN+ M &$\longrightarrow$ &C+N+M &3.32$\times 10^{-10}$& 0& 74989.0 & NIST\\
TF21&SiS+ M &$\longrightarrow$ &Si+S+M &4.40$\times 10^{-10}$& 0& 98600.0 & E as TF13\\
TF22&SiN+ M &$\longrightarrow$ &Si+N+M &3.32$\times 10^{-10}$& 0& 74989.0& E as TF20\\
TF23&S$_2$+ M& $\longrightarrow$ &S+S+M &7.95$\times 10^{-11}$& 0& 38749.0 & NIST\\
TF24&N$_2$+ M& $\longrightarrow$ &N+N+M &2.52$\times 10^{-7}$& -1.6& 57005.4 & NIST\\
\tableline
\multicolumn{8}{c}{\bf NEUTRAL-NEUTRAL}  \\
\tableline
NN1&H + OH&$\longrightarrow$ &H$_2$+ O&6.86$\times 10^{-14}$& 2.8& 1949.5& NIST\\
NN2&H + H$_2$O&$\longrightarrow$ &OH+ H$_2$&6.82$\times 10^{-12}$& 1.6& 9720& NIST\\
NN3&H + O$_2$&$\longrightarrow$ &OH+ O&6.73$\times 10^{-10}$& -0.6& 8151.0& NIST\\
NN4&H + CH &$\longrightarrow$ &C + H$_2$ &1.31$\times 10^{-10}$& -1.6& 80.0& NIST\\
NN5&H + C$_2$&$\longrightarrow$ &C+CH &4.67$\times 10^{-10}$& 0.5& 30450.0& UDFA06\\
NN6&H + CH$2$ &$\longrightarrow$ &CH + H$_2$ &1.00$\times 10^{-11}$& 0& -899.6& NIST\\
NN7&H + C$_2$H &$\longrightarrow$ &C$_2$ + H$_2$ &5.99$\times 10^{-11}$& 0& 11800.0& NIST\\
NN8&H + C$_2$H$_2$ &$\longrightarrow$& C$_2$H + H$_2$ &1$\times 10^{-10}$& 0& 14000.0& NIST\\
NN9&H + HC$_2$O&$\longrightarrow$ &O+C$_2$H$_2$ &2.51$\times 10^{-10}$& 0& 0& NIST\\
NN10&H + HC$_2$O&$\longrightarrow$ &CO+CH$_2$ &4.98$\times 10^{-11}$& 0& 0& NIST\\
NN11&H + HCN&$\longrightarrow$ &CN+ H$_2$&6.19$\times 10^{-10}$& 0& 12499.0& NIST\\
NN12&H + CO&$\longrightarrow$ &C+OH &1.1$\times 10^{-10}$& 0.5& 77700.0& UDFA06\\
NN13&H + CO&$\longrightarrow$ &CO+H &1.0$\times 10^{-16}$& 0& 0& E\\
NN14&H + CO$_2$&$\longrightarrow$& CO+OH &2.51$\times 10^{-10}$& 0& 13349.0& NIST\\
NN15&H + CS&$\longrightarrow$ &CH+S &1.2$\times 10^{-11}$& 0.6& 5880.0& UDFA06\\
NN16&H + SiH&$\longrightarrow$ &Si + H$_2$&2.00$\times 10^{-11}$& 0& 0& WC98\\
NN17&H + SiO&$\longrightarrow$ &SiH + O&1.00$\times 10^{-16}$& 0& 0& E as NN13\\
NN18&H + SiO&$\longrightarrow$ &Si + OH$_2$&1.00$\times 10^{-16}$& 0& 0& NN13\\
NN19&H + SiS&$\longrightarrow$ &SiH+ S&1.00$\times 10^{-16}$& 0& 0& NN13\\
NN20&H + SO&$\longrightarrow$ &OH+ S&5.90$\times 10^{-10}$& -0.3& 11100& UDFA06\\
NN21&H + SO&$\longrightarrow$ &OH+ S&5.90$\times 10^{-10}$& -0.3& 11100& UDFA06\\
NN22&H + NO&$\longrightarrow$ &OH+ N&2.82$\times 10^{-10}$& 0& 24560.0& NIST\\
NN23&O + H$_2$&$\longrightarrow$ &OH+H&3.39$\times 10^{-13}$& 2.7& 3159.3& NIST\\
NN24&O + OH&$\longrightarrow$ &O$_2$+ H&4.55$\times 10^{-12}$& 0.4& -371.6& NIST\\
NN25&O + H$_2$O&$\longrightarrow$ &OH+ OH&1.85$\times 10^{-11}$& 1& 8750& NIST\\
NN26&O + CH&$\longrightarrow$ &OH+ C&2.52$\times 10^{-11}$& 0& 2381.0& NIST\\
NN27&O+ CH&$\longrightarrow$ &CO+ H&6.59$\times 10^{-11}$& 0& 0& NIST\\
NN28&O+ CH$_2$&$\longrightarrow$ &CO+ H$_2$&1.33$\times 10^{-11}$& 0& 0& RH01\\
NN29&O+ CH$_2$&$\longrightarrow$ &CO+ H+H&1.33$\times 10^{-10}$& 0& 0& NIST\\
NN30&O + C$_2$H&$\longrightarrow$ &CO+ CH&1.69$\times 10^{-11}$& 0& 0& NIST\\
NN31&O+ C$_2$H$_2$&$\longrightarrow$ &CO+ CH$_2$&2.66$\times 10^{-10}$& 0& 1980.1& NIST\\
NN32&O+ C$_2$H$_2$&$\longrightarrow$ &HC$_2$O+ H$_2$&1.19$\times 10^{-12}$& 2.0& 956.1& NIST\\
NN33&O+ HC$_2$O&$\longrightarrow$ &C$_2$H+ O$_2$&1.66$\times 10^{-10}$& 0& 0& NIST\\
NN34&O+ HC$_2$O&$\longrightarrow$ &CO$_2$H+ CH&1.99$\times 10^{-12}$& 0& 0& NIST\\
NN35&O + HCN&$\longrightarrow$ &OH+ CN&1.43$\times 10^{-12}$& 1.5& 3799.1& NIST\\
NN36&O + SiH&$\longrightarrow$ &SiO+ H&1.00$\times 10^{-10}$& 0& 0& UDFA06\\
NN37&O + CO&$\longrightarrow$ &O$_2$+ C&1.00$\times 10^{-16}$& 0& 0& E as NN13\\
NN38&O + CO$_2$&$\longrightarrow$ &CO+ O$_2$&2.81$\times 10^{-11}$& 0& 24458.0& NIST\\
NN39&O + C$_2$&$\longrightarrow$ &CO+ C&5.99$\times 10^{-10}$& 0& 0&NIST\\
NN40&O + CS&$\longrightarrow$ &CO+ S&5.00$\times 10^{-11}$& 0& 0& NIST\\
NN41&O + CS&$\longrightarrow$ &SO+ C&4.68$\times 10^{-11}$& 0.5& 28940.0& UDFA06\\
NN42&O + CN&$\longrightarrow$ &CO+ N&1.45$\times 10^{-10}$& -0.18& 0& NIST\\
NN43&O + SiO&$\longrightarrow$ &O$_2$+ Si&1.00$\times 10^{-16}$& 0& 0& E as NN13\\
NN44&O + SiN&$\longrightarrow$ &SiO+ N&6.64$\times 10^{-11}$& 0& 0& NIST\\
NN45&O + SiN&$\longrightarrow$ &NO+ Si&5.00$\times 10^{-11}$& 0& 0& UDFA06\\
NN46&O + SO&$\longrightarrow$ &O$_2$+ S&2.13$\times 10^{-13}$& 1.5& 2536.4& NIST\\
NN47&O + S$_2$S&$\longrightarrow$ &SO+ S&1.70$\times 10^{-11}$& 0& 0& UDFA06\\
NN48&O + NO&$\longrightarrow$ &O$_2$+ N&2.74$\times 10^{-11}$& 0& 21286.0& NIST\\
NN49&O + N$_2$&$\longrightarrow$ &NO+ N&3.01$\times 10^{-10}$& 1.0& 38244.0& NIST\\
NN50&O + SiN&$\longrightarrow$ &SiO+ N&6.64$\times 10^{-11}$& 0& 0& NIST\\
NN51&C + H$_2$&$\longrightarrow$ &CH+ H&6.64$\times 10^{-10}$& 0& 11699.2& NIST\\
NN52&C + OH&$\longrightarrow$ &CO+ H&1.10$\times 10^{-11}$& 0& 0& UDFA06\\
NN53&C + OH&$\longrightarrow$ &CH+ O&2.25$\times 10^{-11}$& 0.5& 14800.0& UDFA06\\
NN54&C + H$_2$O&$\longrightarrow$ &CH+ OH&1.20$\times 10^{-12}$& 0& 19776.0& NIST\\
NN55&C + CH$_2$&$\longrightarrow$ &CH+ CH&2.69$\times 10^{-12}$& 0.5& 0& NIST\\
NN56&C + O$_2$&$\longrightarrow$ &CO+ O&2.46$\times 10^{-12}$& 1.5& -613.0& UDFA06\\
NN57&C + CO&$\longrightarrow$ &C$_2$ + O&5.43$\times 10^{-7}$& -1.5& 57200.0& Hanson 1973\\
NN58&C + CO$_2$&$\longrightarrow$ &CO+ CO&1.10$\times 10^{-15}$& 0& 0& NIST\\
NN59&C + SiO&$\longrightarrow$ &CO+ Si&1.00$\times 10^{-16}$& 0& 0& E as NN13\\
NN60&C + SO&$\longrightarrow$ &CO+ S&3.50$\times 10^{-11}$& 0& 0& UDFA06\\
NN61&C + SO&$\longrightarrow$ &CS+ O&3.50$\times 10^{-11}$& 0& 0& UDFA06\\
NN62&C + S$_2$&$\longrightarrow$ &CS+ S&7.11$\times 10^{-11}$& 0& 0& UDFA06\\
NN63&C + S$_2$&$\longrightarrow$ &CS+ S&7.11$\times 10^{-11}$& 0& 0& UDFA06\\
NN64&C + NO&$\longrightarrow$ &CO+ N&3.49$\times 10^{-11}$& 0& 0& NIST\\
NN65&C + NO&$\longrightarrow$ &CN+ O&5.57$\times 10^{-11}$& -0.3& 0& NIST\\
NN66&C + N$_2$&$\longrightarrow$ &CN+ N&8.69$\times 10^{-11}$& 0& 22600.0& NIST\\
NN67&Si + H$_2$&$\longrightarrow$ &SiH+ H&6.64$\times 10^{-10}$& 0& 11699.0& E as NN51\\
NN68&Si + OH&$\longrightarrow$ &SiO+ H&3.32$\times 10^{-10}$& 0& 0& NIST\\
NN69&Si + O$_2$&$\longrightarrow$ &SiO+ O&1.72$\times 10^{-10}$& -0.5& 16.8& NIST\\
NN70&Si + CO&$\longrightarrow$ &SiO+ C&1.30$\times 10^{-9}$& 0&34516.0& NIST\\
NN71&Si + CO$_2$&$\longrightarrow$ &SiO+ CO&9.96$\times 10^{-10}$&0& 9419.1& NIST\\
NN72&Si + S$_2$&$\longrightarrow$ &SiS+ S&7.00$\times 10^{-11}$&0& 0& E as NN62\\
NN73&Si + NO&$\longrightarrow$ &SiO+ N&5.31$\times 10^{-11}$&0& 1779.9& NIST\\
NN74&Si + O$_2$&$\longrightarrow$ &SiO+ O&1.72$\times 10^{-10}$& -0.5& 16.8& NIST\\
NN75&S + OH&$\longrightarrow$ &SO+ H&6.59$\times 10^{-11}$& 0& 0& NIST\\
NN76&S + O$_2$&$\longrightarrow$ &SO+ O&1.39$\times 10^{-12}$& 0.5& 15.0& NIST\\
NN77&S + CH&$\longrightarrow$ &CS+ H&6.59$\times 10^{-11}$& 0& 0& E as NN27\\
NN78&S + CO&$\longrightarrow$ &SO+ C&1.00$\times 10^{-16}$& 0& 0& E as NN13\\
NN79&S + CO&$\longrightarrow$ &CS+ O&1.00$\times 10^{-16}$& 0& 0& E as NN13\\
NN80&S + CS&$\longrightarrow$ &S$_2$+ C&1.73$\times 10^{-11}$& 0.5& 11500& E as NN83\\
NN81&S + CN&$\longrightarrow$ &CS+ N&1.45$\times 10^{-10}$& -0.2& 0& E as NN42\\
NN82&S + SiS&$\longrightarrow$ &S$_2$+ Si&4.68$\times 10^{-11}$& 0.5& 22&E as NN44\\
NN83&S + SO&$\longrightarrow$ &S$_2$+ O&1.73$\times 10^{-11}$& 0.5& 11500.0& UDFA06\\
NN84&S + NO&$\longrightarrow$ &SO+ N&1.75$\times 10^{-10}$& 0& 20200.0& UDFA06\\
NN85&N + OH&$\longrightarrow$ &NO+ H&4.00$\times 10^{-11}$& 0& 0& NIST\\
NN86&N + HCN&$\longrightarrow$ &N$_2$+ CH&3.00$\times 10^{-10}$& 0& 0& E as NN91\\
NN87&N + O$_2$&$\longrightarrow$ &NO+ O&3.44$\times 10^{-12}$& 1.2& 4000& NIST\\
NN88&N + CO&$\longrightarrow$ &NO+ C&3.84$\times 10^{-9}$& 0& 35959.0& NIST\\
NN89&N + CO&$\longrightarrow$ &CN+ O&3.83$\times 10^{-9}$& 0& 35990.0& NIST\\
NN90&N + CO$_2$&$\longrightarrow$ &NO+ CO&3.20$\times 10^{-13}$& 0& 1710.5& NIST\\
NN91&N + CN&$\longrightarrow$ &N$_2$+ C&3.00$\times 10^{-10}$& 0& 0& NIST\\
NN92&N + CS&$\longrightarrow$ &CN+ S&3.80$\times 10^{-11}$& 0.5& 1160.0& UDFA06\\
NN93&N + SiO&$\longrightarrow$ &SiN+ O&3.84$\times 10^{-9}$& 0& 35959.0& E as NN88\\
NN94&N + SiO&$\longrightarrow$ &NO+ Si&1.00$\times 10^{-16}$& 0& 0& E as NN13\\
NN94&N + SiS&$\longrightarrow$ &SiN+ S&3.80$\times 10^{-11}$& 0& 1160.0& E as NN92\\
NN94&N + SO&$\longrightarrow$ &NO+ S&1.73$\times 10^{-11}$& 0.5& 750& UDFA06\\
NN95&N + NO&$\longrightarrow$ &N$_2$+ O&7.11$\times 10^{-11}$& 0& 786.1& NIST\\
NN96&H$_2$ + OH&$\longrightarrow$ &H$_2$O+ H&2.97$\times 10^{-12}$& 1.2& 2370.4& NIST\\
NN97&H$_2$ + CH&$\longrightarrow$ &CH$_2$+ H&1.48$\times 10^{-11}$& 1.8& 839.4& NIST\\
NN98&H$_2$ + C$_2$H&$\longrightarrow$ &C$_2$H$_2$+ H&8.95$\times 10^{-13}$& 2.6&129.9& NIST\\
NN99&H$_2$ + CO$_2$&$\longrightarrow$ &CH$_2$+ O$_2$&1.00$\times 10^{-16}$& 0& 0& E as NN13\\
NN100&H$_2$ + CO&$\longrightarrow$ &CH$_2$+ O&1.00$\times 10^{-33}$& 0& 0& E\\
NN101&H$_2$ + O$_2$&$\longrightarrow$ &OH+ OH&3.16$\times 10^{-10}$& 0& 21890.0& NIST\\
NN102&H$_2$ + C$_2$&$\longrightarrow$ &C$_2$H+ H&1.10$\times 10^{-10}$& 0& 4000.0& NIST\\
NN103&H$_2$ + CN&$\longrightarrow$ &HCN+ H&5.65$\times 10^{-13}$& 2.4& 1129.9& NIST\\
NN104&OH + CH&$\longrightarrow$ &H$_2$O+ C&1.00$\times 10^{-16}$& 0& 0& E as NN13\\
NN109&OH + C$_2$H$_2$&$\longrightarrow$ &HC$_2$O+ H$_2$&1.91$\times 10^{-13}$& 0& -0& NIST\\
NN105&OH + OH&$\longrightarrow$ &H$_2$+ O$_2$&1.65$\times 10^{-12}$& 1.1& 6013.2& UDFA06\\
NN106&OH + OH&$\longrightarrow$ &H$_2$O+ O$_2$&1.65$\times 10^{-12}$& 1.1& 50.0& UDFA06\\
NN107&OH + CO&$\longrightarrow$ &H+ CO$_2$&3.75$\times 10^{-14}$& 1.6& -401.9& NIST\\
NN108&OH + CO&$\longrightarrow$ &CH+ O$_2$&1.00$\times 10^{-16}$& 0& 0& E as NN13\\
NN110&OH + CN&$\longrightarrow$ &H+ CO$_2$&3.75$\times 10^{-14}$& 1.6& -401.9& NIST\\
NN111&OH + HCN&$\longrightarrow$ &H$_2$O+CN&2.41$\times 10^{-11}$& 0& 5499.7& NIST\\
NN112&O$_2$ + CH&$\longrightarrow$ &CO+ OH&8.00$\times 10^{-11}$& 0& 0& NIST\\
NN113&O$_2$ + CH$_2$&$\longrightarrow$ &H$_2$+ CO$_2$&1.33$\times 10^{-11}$& 0& 0& RH01\\
NN114&O$_2$ + CH$_2$&$\longrightarrow$ &H$_2$O+ CO&2.54$\times 10^{-10}$& -3.3& 1443.0& RH01\\
NN115&O$_2$ + CH$_2$&$\longrightarrow$ &CO$_2$+ H+H&1.33$\times 10^{-11}$& 0& 0& RH01\\
NN116&O$_2$ + CH$_2$&$\longrightarrow$ &CO+ OH+H&1.33$\times 10^{-11}$& 0& 0& RH01\\
NN117&O$_2$ + C$_2$H&$\longrightarrow$ &HC$_2$O+ O&1.00$\times 10^{-12}$& 0& 0& NIST\\
NN118&O$_2$ + CO&$\longrightarrow$ &CO$_2$+ O$_2$&4.20$\times 10^{-12}$& 0& 24053.0& NIST\\
NN119&O$_2$ + N$_2$&$\longrightarrow$ &NO+ NO&1.00$\times 10^{-16}$& 0& 0& E as NN13\\
NN120&CH$_2$ + CH$_2$&$\longrightarrow$ &C$_2$H$_2$+H+H&3.32$\times 10^{-10}$& 0& 5229.8& NIST\\
NN121&CH$_2$ + CH$_2$&$\longrightarrow$ &C$_2$H$_2$+H$_2$&2.62$\times 10^{-9}$& 0& 6009.6& NIST\\
NN122&CO + CH&$\longrightarrow$ &C$_2$H+O&1.00$\times 10^{-16}$& 0& 0& E as NN13\\
NN123&CO + CH$_2$&$\longrightarrow$ &C$_2$H$_2$+O&1.00$\times 10^{-16}$& 0& 0& NIST\\
NN124&CO + SiO&$\longrightarrow$ &CO$_2$+Si&1.00$\times 10^{-16}$& 0& 0& E as NN13\\
NN125&CO + NO&$\longrightarrow$ &CO$_2$+N&1.00$\times 10^{-16}$& 0& 0& E as NN13\\
NN126&CN + H$_2$O&$\longrightarrow$ &HCN+ OH&1.25$\times 10^{-11}$& 0& 3719.0& NIST\\
NN127&NO + NO&$\longrightarrow$ &N$_2$+ O$_2$&2.51$\times 10^{-11}$& 0& 30653.0& UDFA06\\
\tableline
\multicolumn{8}{c}{\bf RADIATION ASSOCIATION}  \\
\tableline
RA1&H + H$^+$&$\longrightarrow$ &HeH$^+$ + h$\nu$&5.26$\times 10^{-20}$&-0.5& 0& UDFA06\\
RA2&He + H$^+$&$\longrightarrow$ &H$_2^+$ + h$\nu$&5.13$\times 10^{-19}$&1.8& 0& UDFA06\\
RA3&O + O&$\longrightarrow$ &O$_2$ + h$\nu$&1.00$\times 10^{-19}$& 0& 0& Babb 1995\\
RA4&O + C&$\longrightarrow$ &CO + h$\nu$&1.58$\times 10^{-17}$& 0.3& 1297.4& Dalgarno 1990\\
RA5&O + C$^+$&$\longrightarrow$ &CO$^+$ + h$\nu$&3.14$\times 10^{-18}$& -0.1& 68.0& Dalgarno 1990\\
RA6&O + Si&$\longrightarrow$ &SiO + h$\nu$&5.52$\times 10^{-18}$& 0.3& 0& Andreazza 1995\\
RA7&O + Si$^+$&$\longrightarrow$ &SiO$^+$ + h$\nu$&5.50$\times 10^{-18}$& 0& 0& Andreazza 1995\\
RA8&O + S&$\longrightarrow$ &SO + h$\nu$&1.11$\times 10^{-19}$& 0.3& 1297.9& Andreazza 2005\\
RA9&C + O$^+$&$\longrightarrow$ &CO$^+$ + h$\nu$&3.14$\times 10^{-18}$& -0.1& 68.0& UDFA06\\
RA10&C + C&$\longrightarrow$ &C$_2$ + h$\nu$&4.36$\times 10^{-18}$& 0.3& 161.3& Singh 2000\\
RA11&C + C$_2$&$\longrightarrow$ &C$_3$ + h$\nu$&1.00$\times 10^{-17}$& 0& 0& Clayton 1999\\
RA12&C + S&$\longrightarrow$ &CS + h$\nu$&4.36$\times 10^{-19}$& 0.2& 0& Andreazza 2005\\
RA13&C + N&$\longrightarrow$ &CN + h$\nu$&7.87$\times 10^{-19}$& 0& 96.0& Singh 2000\\
RA14&Si + S&$\longrightarrow$ &CS + h$\nu$&1.05$\times 10^{-19}$& 0.3& 66.1& Andreazza 2007\\
RA15&S + S&$\longrightarrow$ &S$_2$ + h$\nu$&1.38$\times 10^{-19}$& 0.3& -78.8& Andreazza 2005\\
\tableline
\multicolumn{8}{c}{\bf ION-MOLECULE}  \\
\tableline
IM1&H$_2^+$ + He&$\longrightarrow$ &HeH$^+$ + H&1.30$\times 10^{-10}$& 0& 0& UDFA06\\
IM2&H$_2^+$ + H$_2$&$\longrightarrow$ &H$_3^+$ + H&2.08$\times 10^{-9}$& 0& 0& UDFA06\\
IM3&H$_3^+$ + O&$\longrightarrow$ &OH$^+$ + H$_2$ &8.40$\times 10^{-10}$& 0& 0& UDFA06\\
IM4&H$_3^+$ + Mg&$\longrightarrow$ &Mg$^+$ + H$_2$ + H&1.00$\times 10^{-9}$& 0& 0& UDFA06\\
IM5&H$_3^+$ + Fe&$\longrightarrow$ &fe$^+$ + H$_2$ + H&4.90$\times 10^{-9}$& 0& 0& UDFA06\\
IM6&HeH$^+$ + H&$\longrightarrow$ &H$_2^+$ + He&9.10$\times 10^{-10}$& 0& 0& UDFA06\\
IM7&HeH$^+$ + H$_2$&$\longrightarrow$ &H$_3^+$ + He&1.50$\times 10^{-9}$& 0& 0& UDFA06\\
IM8&He$^+$ + H$_2$&$\longrightarrow$ &He + H$^+$ + H&1.10$\times 10^{-9}$& 0& 0& UDFA06\\
IM9&He$^+$ + OH&$\longrightarrow$ &O$^+$ + H + He&1.10$\times 10^{-9}$& 0& 0& UDFA06\\
IM10&He$^+$ + OH&$\longrightarrow$ &O+ H$^+$ + He&1.10$\times 10^{-9}$& 0& 0& UDFA06\\
IM11&He$^+$ + H$_2$O&$\longrightarrow$ &OH + H$^+$ + He&2.04$\times 10^{-10}$& 0& 0& UDFA06\\
IM12&He$^+$ + O$_2$&$\longrightarrow$ &O$^+$ + O + He&1.10$\times 10^{-9}$& 0& 0& UDFA06\\
IM13&He$^+$ + CO&$\longrightarrow$ &C$^+$ + O + He&1.40$\times 10^{-9}$& -0.5& 0& UDFA06\\
IM14&He$^+$ + CO$_2$&$\longrightarrow$ &CO$^+$ + O + He&8.70$\times 10^{-10}$& 0& 0& UDFA06\\
IM15&He$^+$ + C$_2$&$\longrightarrow$ &C$^+$ + C + He&1.60$\times 10^{-9}$& 0& 0& UDFA06\\
IM16&He$^+$ + CH$_2$&$\longrightarrow$ &C$^+$ + H$_2$ + He&7.50$\times 10^{-10}$& 0& 0& UDFA06\\
IM17&He$^+$ + C$_2$H$_2$&$\longrightarrow$ &C$_2^+$ + H$_2$ + He&1.61$\times 10^{-9}$& 0& 0& UDFA06\\
IM18&He$^+$ + SiO&$\longrightarrow$ &O$^+$ + Si+ He&8.60$\times 10^{-10}$& 0& 0& UDFA06\\
IM19&He$^+$ + SiO&$\longrightarrow$ &Si$^+$ + O + He&8.60$\times 10^{-10}$& 0& 0& UDFA06\\
IM20&C$^+$ + O$^2$&$\longrightarrow$ &CO$^+$ + O&3.80$\times 10^{-10}$& 0& 0& UDFA06\\
IM21&C$^+$ + O$^2$&$\longrightarrow$ &O$^+$ + CO&6.20$\times 10^{-10}$& 0& 0& UDFA06\\
IM22&C$^+$ + CO$^2$&$\longrightarrow$ &CO$^+$ + CO&1.10$\times 10^{-9}$& 0& 0& UDFA06\\
IM23&C$^+$ + SiO&$\longrightarrow$ &Si$^+$ + CO&5.40$\times 10^{-10}$& 0& 0& UDFA06\\
IM24&C$^+$ + SO&$\longrightarrow$ &S$^+$ + CO&2.60$\times 10^{-10}$& 0& 0& UDFA06\\
IM25&C$^+$ + SO&$\longrightarrow$ &CO$^+$ + S&2.60$\times 10^{-10}$& 0& 0& UDFA06\\
IM26&Si$^+$ + OH&$\longrightarrow$ &SiO$^+$ +H&6.30$\times 10^{-10}$& 0& 0& UDFA06\\
IM27&S$^+$ + OH&$\longrightarrow$ &SO$^+$ + H&6.10$\times 10^{-10}$& 0& 0& UDFA06\\
IM28&S$^+$ + O$_2$&$\longrightarrow$ &SO$^+$ + O&1.50$\times 10^{-11}$& 0& 0& UDFA06\\
IM29&C$_2$$^+$ + O&$\longrightarrow$ &CO$^+$ + C&3.10$\times 10^{-10}$& 0& 0& UDFA06\\
IM30&SiO$^+$ + O&$\longrightarrow$ &Si$^+$ +O$_2$&2.00$\times 10^{-10}$& 0& 0& UDFA06\\
IM31&SiO$^+$ + C&$\longrightarrow$ &Si$^+$ + CO&1.00$\times 10^{-9}$& 0& 0& UDFA06\\
IM32&SiO$^+$ + CO&$\longrightarrow$ &Si$^+$ + CO$_2$&7.90$\times 10^{-10}$& 0& 0& UDFA06\\
IM33&SiO$^+$ + S&$\longrightarrow$ &Si$^+$ + SO$_2$&1.00$\times 10^{-9}$& 0& 0& UDFA06\\
IM34&SiO$^+$ + N&$\longrightarrow$ &Si$^+$ + NO&2.10$\times 10^{-10}$& 0& 0& UDFA06\\
\tableline
\multicolumn{8}{c}{\bf CHARGE EXCHANGE}  \\
\tableline
CE1&H + H$_2^+$&$\longrightarrow$ &H$^+$ + H$_2$&6.40$\times 10^{-10}$& 0& 0& UDFA06\\
CE2&O + C$_2^+$&$\longrightarrow$ &CO$^+$ + C&3.10$\times 10^{-10}$& 0& 0& UDFA06\\
CE3&O + CO$^+$&$\longrightarrow$ &O$^+$ + CO&1.40$\times 10^{-10}$& 0.5& 0& UDFA06\\
CE4&O + C$_2^+$&$\longrightarrow$ &CO$^+$ + C&3.10$\times 10^{-10}$& 0& 0& UDFA06\\
CE5&C + CO$^+$&$\longrightarrow$ &C$^+$ + CO&1.10$\times 10^{-10}$& 0& 0& UDFA06\\
CE6&C + C$_2^+$&$\longrightarrow$ &C$^+$ + C$_2$&1.10$\times 10^{-10}$& 0& 0& UDFA06\\
CE7&C$_2$ + O$^+$&$\longrightarrow$ &C$_2^+$ + O&4.80$\times 10^{-10}$& 0& 0& UDFA06\\
CE8&CO + O$^+$&$\longrightarrow$ &CO$^+$ + O&4.90$\times 10^{-12}$& 0.5& 4580.0& UDFA06\\
CE9&Si + C$^+$&$\longrightarrow$ &Si$^+$ + C&2.10$\times 10^{-9}$& 0& 0& UDFA06\\
CE10&Si + S$^+$&$\longrightarrow$ &Si$^+$ + S&1.60$\times 10^{-9}$& 0& 0& NIST\\
CE11&SiO + H$^+$&$\longrightarrow$ &SiO$^+$ + H&3.30$\times 10^{-9}$& 0& 0& UDFA06\\
CE12&S + C$^+$&$\longrightarrow$ &S$^+$ + C&1.50$\times 10^{-9}$& 0& 0& NIST\\
CE13&S + CO$^+$&$\longrightarrow$ &S$^+$ + CO&1.10$\times 10^{-9}$& 0& 0& NIST\\
CE14&Fe + O$^+$&$\longrightarrow$ &Fe$^+$ + O&2.90$\times 10^{-9}$& 0& 0& UDFA06\\
CE15&Fe + C$^+$&$\longrightarrow$ &Fe$^+$ + C&2.60$\times 10^{-9}$& 0& 0& UDFA06\\
CE16&Fe + Si$^+$&$\longrightarrow$ &Fe$^+$ + Si&1.90$\times 10^{-9}$& 0& 0& UDFA06\\
CE17&Fe + S$^+$&$\longrightarrow$ &Fe$^+$ + S&1.80$\times 10^{-10}$& 0& 0& NIST\\
CE18&Fe + SO$^+$&$\longrightarrow$ &Fe$^+$ + SO&1.60$\times 10^{-9}$& 0& 0& UDFA06\\
CE19&Mg + H$^+$&$\longrightarrow$ &Mg$^+$ + H&1.10$\times 10^{-9}$& 0& 0& UDFA06\\
CE20&Mg + O$^+$&$\longrightarrow$ &Mg$^+$ + O&1.10$\times 10^{-9}$& 0& 0& UDFA06\\
CE21&Mg + C$^+$&$\longrightarrow$ &Mg$^+$ + C&1.10$\times 10^{-9}$& 0& 0& UDFA06\\
CE22&Mg + Si$^+$&$\longrightarrow$ &Mg$^+$ + Si&2.90$\times 10^{-9}$& 0& 0& UDFA06\\
CE23&Mg +SiO$^+$&$\longrightarrow$ &Mg$^+$ + SiO&1.00$\times 10^{-9}$& 0& 0& UDFA06\\
CE24&Al + O$^+$&$\longrightarrow$ &Al$^+$ + O&2.00$\times 10^{-9}$& 0& 0& UDFA06\\
CE25&Al + C$^+$&$\longrightarrow$ &Al$^+$ + C&2.70$\times 10^{-9}$& 0& 0& UDFA06\\
CE26&Al + CO$^+$&$\longrightarrow$ &Al$^+$ + CO&2.20$\times 10^{-9}$& 0& 0& UDFA06\\
CE27&Al + Si$^+$&$\longrightarrow$ &Al$^+$ + Si&2.70$\times 10^{-9}$& 0& 0& UDFA06\\
CE28&Al + SiO$^+$&$\longrightarrow$ &Al$^+$ + SiO&2.20$\times 10^{-9}$& 0& 0& UDFA06\\
\tableline
\multicolumn{8}{c}{\bf ELECTRONIC RECOMBINATION}  \\
\tableline
ER1&H$^+$ + e$^-$&$\longrightarrow$ &H&3.50$\times 10^{-12}$& -0.7& 0& UDFA06\\
ER2&H$_2^+$ + e$^-$&$\longrightarrow$ &H + H&1.60$\times 10^{-8}$& 1.4& 0& UDFA06\\
ER3&H$_3^+$ + e$^-$&$\longrightarrow$ &H$_2$ + H&2.34$\times 10^{-8}$& -0.5& 0& UDFA06\\
ER4&H$_3^+$ + e$^-$&$\longrightarrow$ &H$_2$ + H&2.34$\times 10^{-8}$& -0.5& 0& UDFA06\\
ER5&C$^+$ + e$^-$&$\longrightarrow$ &C&4.67$\times 10^{-12}$& -0.6& 0& UDFA06\\
ER6&C$_2^+$ + e$^-$&$\longrightarrow$ &C + C &3.00$\times 10^{-7}$& -0.5& 0& UDFA06\\
ER7&CO$^+$ + e$^-$&$\longrightarrow$ &C + O&2.00$\times 10^{-7}$& -0.5& 0& UDFA06\\
ER8&O$^+$ + e$^-$&$\longrightarrow$ &O&3.24$\times 10^{-12}$& -0.7& 0& UDFA06\\
ER9&Si$^+$ + e$^-$&$\longrightarrow$ &Si&4.90$\times 10^{-12}$& -0.6& 0& UDFA06\\
ER10&SiO$^+$ + e$^-$&$\longrightarrow$ &Si + O&2.00$\times 10^{-7}$& -0.5& 0& UDFA06\\
ER11&S$^+$ + e$^-$&$\longrightarrow$ &S&3.90$\times 10^{-12}$& -0.6& 0& UDFA06\\
ER12&SO$^+$ + e$^-$&$\longrightarrow$ &S + O&2.00$\times 10^{-7}$& -0.5& 0& UDFA06\\
ER13&Fe$^+$ + e$^-$&$\longrightarrow$ &Fe&3.70$\times 10^{-12}$& -0.6& 0& UDFA06\\
ER14&Mg$^+$ + e$^-$&$\longrightarrow$ &Mg&2.80$\times 10^{-12}$& -0.9& 0& UDFA06\\
ER15&Al$^+$ + e$^-$&$\longrightarrow$ &Al&3.24$\times 10^{-12}$& -0.7& 0& UDFA06\\
\tableline
\multicolumn{8}{c}{\bf COMPTON ELECTRON DESTRUCTION}  \\
\tableline
CED1&O&$\longrightarrow$ &O$^+$ + e$^-$&1.57$\times 10^{-5}$& 0& 3464.1& See text \\
CED2&C&$\longrightarrow$ &C$^+$ + e$^-$&1.99$\times 10^{-5}$& 0& 3464.1&  \\
CED3&Si&$\longrightarrow$ &Si$^+$ + e$^-$&1.57$\times 10^{-5}$& 0& 3464.1&  \\
CED4&S&$\longrightarrow$ &S$^+$ + e$^-$&1.57$\times 10^{-5}$& 0& 3464.1&  \\
CED5&Fe&$\longrightarrow$ &Fe$^+$ + e$^-$&1.57$\times 10^{-5}$& 0& 3464.1&  \\
CED6&Mg&$\longrightarrow$ &Mg$^+$ + e$^-$&1.57$\times 10^{-5}$& 0& 3464.1&  \\
CED7&Al&$\longrightarrow$ &Al$^+$ + e$^-$&1.57$\times 10^{-5}$& 0& 3464.1&  \\
CED8&H$_2$&$\longrightarrow$ &H + H$^+$ + e$^-$&8.86$\times 10^{-7}$& 0& 3464.1&  \\
CED9&H$_2$&$\longrightarrow$ &H$_2^+$ + e$^-$&1.93$\times 10^{-5}$& 0& 3464.1&  \\
CED10&H$_2$&$\longrightarrow$ &H + H&9.43$\times 10^{-6}$& 0& 3464.1&  \\
CED11&OH&$\longrightarrow$ &H + O$^+$ + e$^-$&9.46$\times 10^{-7}$& 0& 3464.1&  \\
CED12&OH&$\longrightarrow$ &O + H$^+$ + e$^-$&2.94$\times 10^{-6}$& 0& 3464.1&  \\
CED13&OH&$\longrightarrow$ &O + H&5.81$\times 10^{-6}$& 0& 3464.1&  \\
CED14&H$_2$O&$\longrightarrow$ &OH + H$^+$ + e$^-$&2.94$\times 10^{-6}$& 0& 3464.1&  \\
CED15&H$_2$O&$\longrightarrow$ &OH + H&5.81$\times 10^{-6}$& 0& 3464.1&  \\
CED16&CH&$\longrightarrow$ &H + C$^+$ + e$^-$&9.46$\times 10^{-7}$& 0& 3464.1&  \\
CED17&CH&$\longrightarrow$ &C + H$^+$ + e$^-$&2.94$\times 10^{-6}$& 0& 3464.1&  \\
CED18&CH&$\longrightarrow$ &H + C&5.81$\times 10^{-6}$& 0& 3464.1&  \\
CED19&CH$_2$&$\longrightarrow$ &CH + H$^+$ + e$^-$&2.94$\times 10^{-6}$& 0& 3464.1&  \\
CED20&CH$_2$&$\longrightarrow$ &CH + H&5.81$\times 10^{-6}$& 0& 3464.1&  \\
CED21&C$_2$H$_2$&$\longrightarrow$ &C$_2$H + H$^+$ + e$^-$&2.94$\times 10^{-6}$& 0& 3464.1&  \\
CED22&C$_2$H$_2$&$\longrightarrow$ &C$_2$H + H&5.81$\times 10^{-6}$& 0& 3464.1&  \\
CED23&HCO&$\longrightarrow$ &CO + H$^+$ + e$^-$&2.94$\times 10^{-6}$& 0& 3464.1&  \\
CED24&HCO&$\longrightarrow$ &H + CO$^+$ + e$^-$&8.86$\times 10^{-7}$& 0& 3464.1&  \\
CED25&HCO&$\longrightarrow$ &H + CO&5.81$\times 10^{-6}$& 0& 3464.1&  \\
CED26&O$_2$&$\longrightarrow$ &O + O$^+$ + e$^-$&9.46$\times 10^{-7}$& 0& 3464.1&  \\
CED27&O$_2$&$\longrightarrow$ &O + O&5.81$\times 10^{-6}$& 0& 3464.1&  \\
CED28&CO&$\longrightarrow$ &C + O$^+$ + e$^-$&9.46$\times 10^{-7}$& 0& 3464.1&  \\
CED29&CO&$\longrightarrow$ &O + C$^+$ + e$^-$&2.94$\times 10^{-6}$& 0& 3464.1&  \\
CED30&CO&$\longrightarrow$ &CO$^+$ + e$^-$&2.14$\times 10^{-5}$& 0& 3464.1&  \\
CED31&CO&$\longrightarrow$ &C + O&5.81$\times 10^{-7}$& 0& 3464.1&  \\
CED32&CO$_2$&$\longrightarrow$ &CO + O$^+$ + e$^-$&9.46$\times 10^{-7}$& 0& 3464.1&  \\
CED33&CO$_2$&$\longrightarrow$ &O + CO$^+$ + e$^-$&2.94$\times 10^{-6}$& 0& 3464.1&  \\
CED34&CO$_2$&$\longrightarrow$ &CO + O&5.81$\times 10^{-7}$& 0& 3464.1&  \\
CED35&CO$_2$&$\longrightarrow$ &CO + O$^+$ + e$^-$&9.46$\times 10^{-7}$& 0& 3464.1&  \\
CED36&C$_2$&$\longrightarrow$ &C + C$^+$ + e$^-$&9.46$\times 10^{-7}$& 0& 3464.1&  \\
CED37&C$_2$&$\longrightarrow$ &C$_2^+$ + e$^-$&2.14$\times 10^{-5}$& 0& 3464.1&  \\
CED38&C$_2$&$\longrightarrow$ &C + C&5.81$\times 10^{-6}$& 0& 3464.1&  \\
CED39&SiO&$\longrightarrow$ &Si + O$^+$ + e$^-$&1.07$\times 10^{-7}$& 0& 3464.1&  \\
CED40&SiO&$\longrightarrow$ &O + Si$^+$ + e$^-$&3.33$\times 10^{-6}$& 0& 3464.1&  \\
CED41&SiO&$\longrightarrow$ &SiO$^+$ + e$^-$&2.42$\times 10^{-5}$& 0& 3464.1&  \\
CED42&SiO&$\longrightarrow$ &O + Si&6.58$\times 10^{-6}$& 0& 3464.1&  \\
CED43&SiS&$\longrightarrow$ &Si + S$^+$ + e$^-$&2.94$\times 10^{-6}$& 0& 3464.1&  \\
CED44&SiS&$\longrightarrow$ &S + Si$^+$ + e$^-$&2.94$\times 10^{-6}$& 0& 3464.1&  \\
CED45&SiS&$\longrightarrow$ &Si + S&5.81$\times 10^{-6}$& 0& 3464.1&  \\
CED46&SO&$\longrightarrow$ &O + S$^+$ + e$^-$&2.94$\times 10^{-6}$& 0& 3464.1&  \\
CED47&SO&$\longrightarrow$ &S + O$^+$ + e$^-$&9.46$\times 10^{-7}$& 0& 3464.1&  \\
CED48&SO&$\longrightarrow$ &SO$^+$ + e$^-$&2.14$\times 10^{-5}$& 0& 3464.1&  \\
CED49&SO&$\longrightarrow$ &O + S&5.81$\times 10^{-6}$& 0& 3464.1&  \\
CED50&SO&$\longrightarrow$ &O + S$^+$ + e$^-$&2.94$\times 10^{-6}$& 0& 3464.1&  \\
CED51&S$_2$&$\longrightarrow$ &S + S$^+$ + e$^-$&9.46$\times 10^{-7}$& 0& 3464.1&  \\
CED52&S$_2$&$\longrightarrow$ &S + S&5.81$\times 10^{-6}$& 0& 3464.1&  \\
CED53&N$_2$&$\longrightarrow$ &N + N&5.44$\times 10^{-6}$& 0& 3464.1&  \\
\tableline
\multicolumn{8}{c}{\bf UV PHOTODISSOCIATION}  \\
\tableline
PHOT1&H$_2$&$\longrightarrow$ &H+ H&3.80$\times 10^{-5}$& 0& 3464.1& See text \\
PHOT2&O$_2$&$\longrightarrow$ &O + O&3.80$\times 10^{-5}$& 0& 3464.1&  \\
PHOT3&OH&$\longrightarrow$& O + H&3.80$\times 10^{-5}$& 0& 3464.1&  \\
PHOT4&H$_2$O&$\longrightarrow$ &OH + H&3.80$\times 10^{-5}$& 0& 3464.1&  \\
PHOT5&CO&$\longrightarrow$& C + O&3.80$\times 10^{-5}$& 0& 3464.1&  \\
PHOT6&CO$_2$&$\longrightarrow$ &CO + O&3.80$\times 10^{-5}$& 0& 3464.1&  \\
PHOT7&C$_2$&$\longrightarrow$ &C + C&3.80$\times 10^{-5}$& 0& 3464.1&  \\
PHOT8&CO$_2$&$\longrightarrow$ &CO + O&3.80$\times 10^{-5}$& 0& 3464.1&  \\
PHOT9&SiO&$\longrightarrow$ &Si + O&3.80$\times 10^{-5}$& 0& 3464.1&  \\
PHOT10&SO$_2$&$\longrightarrow$& S + O&3.80$\times 10^{-5}$& 0& 3464.1&  \\
PHOT11&N$_2$&$\longrightarrow$ &N + N&3.80$\times 10^{-5}$& 0& 3464.1&  \\
\enddata
\tablenotetext{a}{NIST is the NIST chemical kinetics database. UDFA06 is by Woodall et al. (2007). Other references are listed in the bibliography. 'E' means 'estimated'. }
\end{deluxetable}
\clearpage

\end{document}